\documentclass{article}
\usepackage{booktabs,tabularx,graphicx}
\usepackage[margin=1in]{geometry}
\usepackage{times}
\usepackage{amsmath}
\usepackage{hyperref}
\usepackage{setspace}
\usepackage{float}
\usepackage{adjustbox}
\usepackage{lscape}
\usepackage{pdflscape}
\usepackage{longtable}
\usepackage{multirow}
\usepackage{caption}
\usepackage{tabularx}%
\usepackage{placeins}
\newcolumntype{Y}{>{\centering\arraybackslash}X}

\title{Statin Recommendations among US Adults with the \\2026 Dyslipidemia Guidelines}

\author{
  James A.\ Diao$^{1,2*}$,
  Thomas A.\ Buckley$^{1*}$,
  Andrew Z.\ Zhou$^{1}$,
  Smaraki Dash$^{2,4}$, \\
  Rishi K.\ Wadhera$^{3,4}$, and
  Arjun K.\ Manrai$^{1**}$
}
\date{
  \vspace{0.5em}
  {\normalsize
  $^{1}$Department of Biomedical Informatics, Harvard Medical School, Boston, MA\\[0.1em]
  $^{2}$Department of Medicine, Brigham and Women's Hospital, Boston, MA\\[0.1em]
  $^{3}$Division of Cardiology, Beth Israel Deaconess Medical Center, Boston, MA\\[0.1em]
  $^{4}$Richard A.\ and Susan F.\ Smith Center for Outcomes Research, Beth Israel Deaconess Medical Center, Boston, MA\\[0.2em]
  $^{*}$Co-first authors\quad $^{**}$Corresponding author: \href{mailto:Arjun_Manrai@hms.harvard.edu}{Arjun\_Manrai@hms.harvard.edu}
  }
}

\begin{document}
\newgeometry{margin=0.85in}
\maketitle

\noindent\textbf{Key Points}

\smallskip
\noindent\textbf{Question}\quad How will new dyslipidemia guidelines change statin recommendations in the US?

\smallskip
\noindent\textbf{Findings}\quad In this cross-sectional analysis of nationally representative data, the 2026 guidelines at the class~1 threshold reduced statin recommendations among US adults by an estimated 3.0 million, with larger reductions among Black, male, and middle-aged adults. However, the class~2 threshold increased statin recommendations by an estimated 20.8 million, driven primarily by elevated 30-year risk.

\smallskip
\noindent\textbf{Meaning}\quad The net effect of the 2026 dyslipidemia guidelines on the number of US adults recommended for statin therapy depends critically on threshold class.

\bigskip
\hrule
\bigskip

\section*{Abstract}

\noindent\textbf{Importance}\quad The 2026 multisociety dyslipidemia guideline recommended the Predicting Risk of cardiovascular disease EVENTs (PREVENT) equations in place of the pooled cohort equations (PCE), introduced 30-year risk assessment as a new treatment pathway, and lowered risk-based treatment thresholds. The net population impact of these concurrent changes on statin recommendations is unknown.

\smallskip
\noindent\textbf{Objective}\quad To estimate changes in statin recommendations under 2026 PREVENT-based dyslipidemia guidelines compared with 2018 PCE-based guidelines.

\smallskip
\noindent\textbf{Design, Setting, and Participants}\quad Cross-sectional analysis of pooled data from the National Health and Nutrition Examination Survey (NHANES), spanning 2011--2023 and comprising 24{,}199 participants aged 30--79 years.

\smallskip
\noindent\textbf{Main Outcomes and Measures}\quad Number and proportion of US adults receiving or recommended for statin therapy.

\smallskip
\noindent\textbf{Results}\quad At the class~1 threshold (PCE $\geq$7.5\% vs PREVENT $\geq$5\%), the number of US adults receiving or recommended for statin therapy decreased by an estimated 3.0 million (95\% confidence interval [CI], 2.3 million to 3.6 million), with larger reductions among Black adults ($-$4.2 percentage points [pp]), men ($-$4.0pp), and adults aged 50--69 years ($-$5.6pp). At the class~2 threshold---which additionally recommends statins for adults aged 30--59 years based on 30-year risk---the number of adults recommended increased by an estimated 20.8 million (95\% CI, 19.6 million to 22.0 million), or +11.6pp. The increase was largest among adults aged 50--59 years (+19.7pp) and 40--49 years (+14.8pp).

\smallskip
\noindent\textbf{Conclusions and Relevance}\quad The net population impact of the 2026 dyslipidemia guidelines depends critically on which recommendation class is applied. At the class~1 threshold, statin recommendations decreased modestly; at the class~2 threshold, inclusion of 30-year ASCVD risk assessment substantially expanded recommendations, particularly among younger adults. These divergent effects underscore the importance of the 30-year risk criterion as a major driver of new eligibility and the need for outcomes and equity monitoring during guideline implementation.

\restoregeometry
\newpage
\section*{Introduction}

The 2026 multisociety dyslipidemia guideline introduces three simultaneous changes to statin recommendations for primary prevention: it replaces the pooled cohort equations (PCE) with the Predicting Risk of cardiovascular disease EVENTs (PREVENT) equations for 10-year atherosclerotic cardiovascular disease (ASCVD) risk estimation, it introduces 30-year ASCVD risk assessment as a new pathway for statin consideration, and it lowers the risk-based thresholds at which statins can or should be discussed with patients \cite{Blumenthal2026}. PREVENT was developed to improve on long-standing PCE calibration concerns \cite{Muntner2014, Rana2016, DeFilippis2017, Yadlowsky2018}, removes race as an input variable, incorporates kidney function as an input variable, and showed improved calibration in some validation cohorts \cite{Khan2024}. Two nationally representative NHANES analyses estimated that substituting PREVENT for PCE while retaining the original treatment thresholds would reduce statin recommendations by 14 million to 17 million adults \cite{Diao2024, Anderson2024}.

The population-level consequences of simultaneously adopting PREVENT, introducing 30-year risk, and lowering treatment thresholds, and how those effects are distributed across demographic groups, remain unquantified. Prior studies evaluated PREVENT at prior thresholds established in 2013 and continued in 2018 \cite{Diao2024, Anderson2024} or examined the equity implications of threshold adjustment without the complete 2026 guideline \cite{Yan2026}, but none have assessed the net effect of all changes together. Because even small shifts in treatment thresholds could translate into millions of adults newly or no longer recommended for preventive therapy \cite{Pencina2014} and because persistent racial, ethnic, and sex disparities in guideline-concordant statin use are well-documented \cite{Frank2023, Kim2023}, a nationally representative analysis of the 2026 guideline is needed.

In this study, we used pooled NHANES data from 2011 through 2023 to estimate reclassifications when using PREVENT-based versus PCE-based risk categories, national projections for changes to statin recommendations under 2026 dyslipidemia guidelines compared with 2018 guidelines, and how these effects differ by age, gender, and race/ethnicity.

\section*{Methods}

\subsection*{Study Design and Population}

We conducted a cross-sectional analysis of pooled data from five NHANES cycles: 2011--2012, 2013--2014, 2015--2016, 2017--March 2020 (prepandemic \cite{Akinbami2022}), and 2021--2023. The study population included all adults aged 30--79 years who completed the mobile examination center (MEC) visit. We excluded pregnant participants. The final analytic sample comprised 24{,}199 adults representing approximately 180.2 million US adults.

\subsection*{Variable Definitions}

Total cholesterol, high-density lipoprotein cholesterol (HDL-C), and triglycerides were measured in serum. Low-density lipoprotein cholesterol (LDL-C) was estimated using the Martin--Hopkins method across all cycles. Estimated glomerular filtration rate (eGFR) was calculated using the Chronic Kidney Disease (CKD) Epidemiology Collaboration (CKD-EPI) 2021 race-free creatinine-based equation, implemented via the \texttt{kidney.epi} R package. Systolic blood pressure (SBP) was obtained from the MEC examination. Smoking status was defined as self-report of having smoked at least 100 cigarettes in one's lifetime and currently smoking every day or some days.

Diabetes was defined as self-reported physician diagnosis, use of glucose-lowering medication, glycohemoglobin (HbA1c) $\geq$6.5\%, or glucose level $\geq$126 mg/dL among participants in the fasting subsample. CKD stage 3--4 was defined as eGFR 15--59 mL/min/1.73m$^2$. Human immunodeficiency virus (HIV) status was ascertained from positive serological testing or antiretroviral medication use. Clinical ASCVD history was defined as self-reported coronary heart disease, myocardial infarction, angina, or stroke. Severe hypercholesterolemia was defined as LDL-C $\geq$190 mg/dL.

In NHANES cycles 2011--2012 through 2017--March 2020, statin use was ascertained from the prescription medication questionnaire, which records generic drug names for all medications used in the past 30 days; other lipid-lowering therapies (ezetimibe, proprotein convertase subtilisin/kexin type-9 [PCSK9] inhibitors, fibrates, niacin) and antihypertensive medications were similarly identified. For the 2021--2023 cycle, in which prescription drug names were not released, statin use was instead ascertained from self-reported use of cholesterol medication. In pre-2021 cycles where both sources were available, self-reported cholesterol medication showed high agreement with prescription-identified statin use (Supplementary Methods).

\subsection*{Risk Score Computation}

PCE 10-year ASCVD risk was computed using the sex- and race-specific Goff 2013 equations for adults aged 40--79 years \cite{Goff2014}; adults aged 30--39 years were not assigned PCE scores because these equations are not validated below age 40. PREVENT 10-year ASCVD risk was computed using the base PREVENT model \cite{Khan2024} for all adults aged 30--79 years; PREVENT incorporates age, sex, total cholesterol, HDL-C, systolic blood pressure, blood pressure treatment, statin use, diabetes, smoking, BMI, and eGFR. PREVENT 30-year ASCVD risk was also computed using the base PREVENT model for all adults aged 30--79 years; this was used for the class~2 recommendation threshold, which recommends statins for adults aged 30--59 years with a 30-year ASCVD risk $\geq$10\%. Continuous risk factor values were capped at each calculator's recommended input ranges to model practical clinical use; a sensitivity analysis was performed to exclude participants with out-of-range values.

\subsection*{Guideline Definitions and Thresholds}

For adults aged 40--75 years, risk-independent indications---diabetes, LDL-C $\geq$190 mg/dL, clinical ASCVD history, CKD stage 3--4, and HIV---conferred automatic statin recommendations among the appropriate age ranges and LDL-C levels under both guidelines regardless of predicted risk. Under 2018 guidelines, risk-based thresholds were applied using PCE for adults aged 40--75 years only. Under 2026 guidelines, PREVENT-based risk thresholds were applied to the full age range of 30--79 years \cite{Khan2024}.

We evaluated two threshold pairings. Our primary analysis used the class~1 threshold of PREVENT $\geq$5\% vs PCE $\geq$7.5\%. Our secondary analysis used the class~2 threshold, which additionally recommends statins for adults with 10-year PREVENT ASCVD risk $\geq$3\% (vs PCE $\geq$5\%), LDL-C 160--189 mg/dL in adults aged 30--59 years, or 30-year PREVENT ASCVD risk $\geq$10\% in adults aged 30--59 years. To isolate the separate effects of the calculator switch and the threshold reduction, we also evaluated the calculator-only change (PREVENT at the prior threshold) and the threshold-only change (PCE at the new threshold).

A participant was classified as ``receiving or recommended'' for statin therapy if they were currently using a statin or would meet criteria for a statin recommendation under the given guideline, following the approach of Diao et al \cite{Diao2024}.

\subsection*{Statistical Analysis}

Analyses were conducted in R version 4.4 using the \texttt{survey} package with NHANES-provided survey weights and design. We estimated weighted population counts, proportions, and 95\% confidence intervals (CIs) for statin recommendations under each guideline. CIs for differences between guidelines were computed using paired within-person contrasts to account for covariance between overlapping eligibility subpopulations. Subgroup analyses were performed by age group (30--39, 40--49, 50--59, 60--69, 70--75, 76--79 years), gender, and race/ethnicity (non-Hispanic [NH] White, NH Black, Hispanic, and NH Asian). Additional subgroup analyses by socioeconomic indicators (poverty-income ratio, insurance status, education) and NHANES cycle are shown in Figure~\ref{fig:sfig_ses}.

We performed several sensitivity analyses, including: (1) adjustment for cholesterol values among participants receiving lipid-lowering therapy to approximate untreated baseline levels (detreatment), (2) exclusion of participants currently receiving statin therapy, (3) exclusion of participants with risk factor inputs outside the recommended ranges for PCE or PREVENT; and (4) reanalysis of the fasting subsample with its associated survey weights.

This study adhered to the Strengthening the Reporting of Observational Studies in Epidemiology (STROBE) reporting guidelines for cross-sectional studies \cite{vonElm2007STROBE}.

\section*{Results}

\subsection*{Study Population}

The analytic sample included 24{,}199 adults aged 30--79 years with medical examination data, representing an estimated 180.2 million US adults when pooled across the entire study period (Table~\ref{tab:table1}). The median age was 52.0 years, 51.5\% were women, and 21.9\% were currently using statins. The prevalence of diabetes was 16.9\%, CKD stage 3--4 was 5.0\%, and clinical ASCVD was 8.9\%. The median 10-year ASCVD risk was 6.0\% with PCE and 2.5\% with PREVENT.

\subsection*{Risk Category Reclassification}

\begin{figure*}[htbp]
\centering
\includegraphics[width=\textwidth]{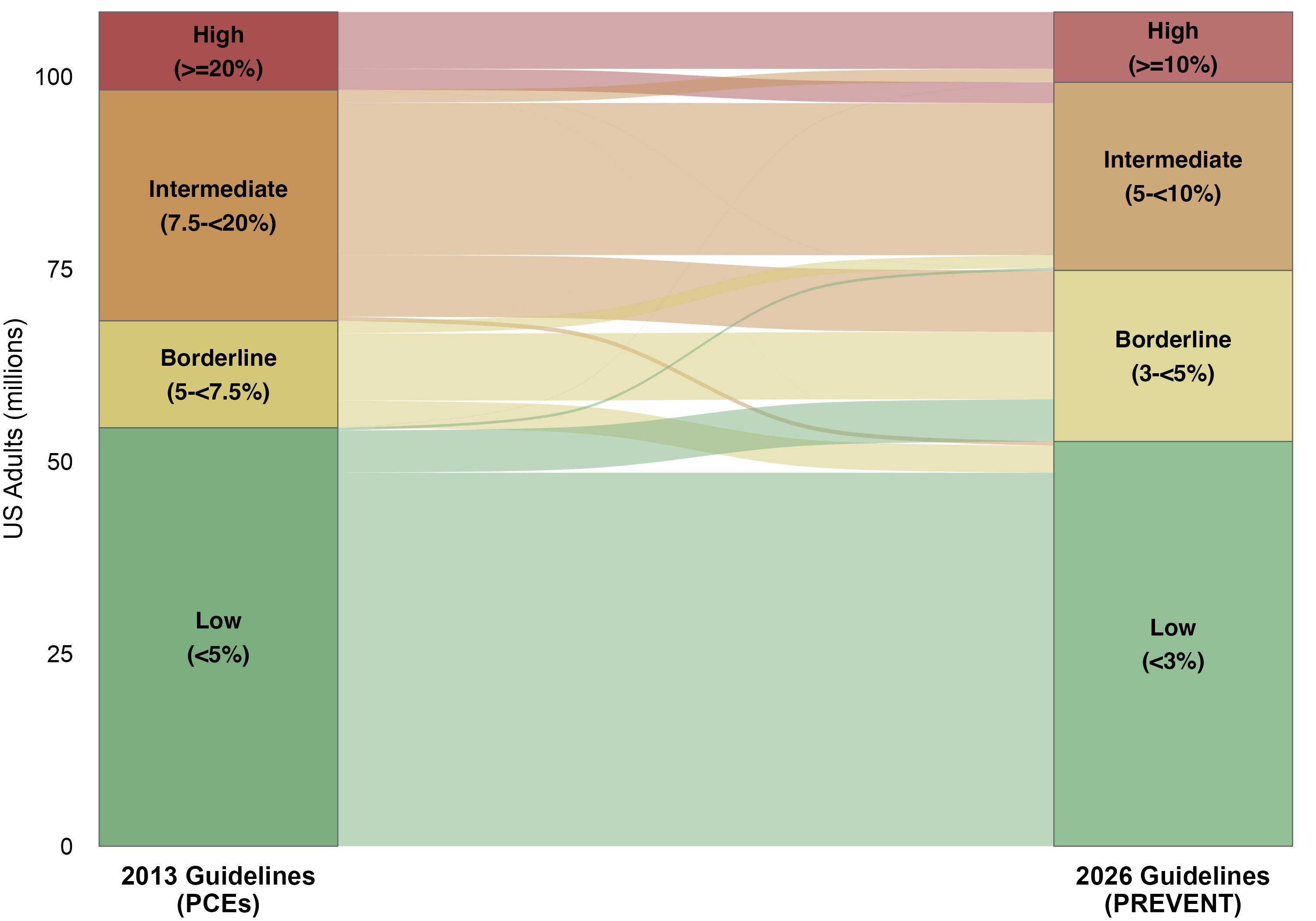}
\caption{\textbf{Risk category reclassification from PCE to PREVENT among US adults aged 40--75 years without clinical ASCVD, NHANES 2011--2023.} Alluvial diagram showing the flow of adults between PCE risk categories (low $<$5\%, borderline 5--$<$7.5\%, intermediate 7.5--$<$20\%, high $\geq$20\%) and PREVENT risk categories (low $<$3\%, borderline 3--$<$5\%, intermediate 5--$<$10\%, high $\geq$10\%). Stratum height is proportional to weighted population count.}
\label{fig:figure1}
\end{figure*}

Figure~\ref{fig:figure1} displays the reclassification of adults aged 40--75 years between PCE risk categories and the corresponding PREVENT categories among adults without clinical ASCVD. Overall, 78.2\% of adults remained in the corresponding risk category, 13.5\% shifted to a lower PREVENT category, and 8.4\% shifted to a higher PREVENT category. Notably, many adults previously in the intermediate category moved to borderline or low categories, reflecting PREVENT's systematically lower risk estimates (Figure~\ref{fig:sfig_scatter}).

\begin{figure*}[htbp]
\centering
\includegraphics[width=\textwidth]{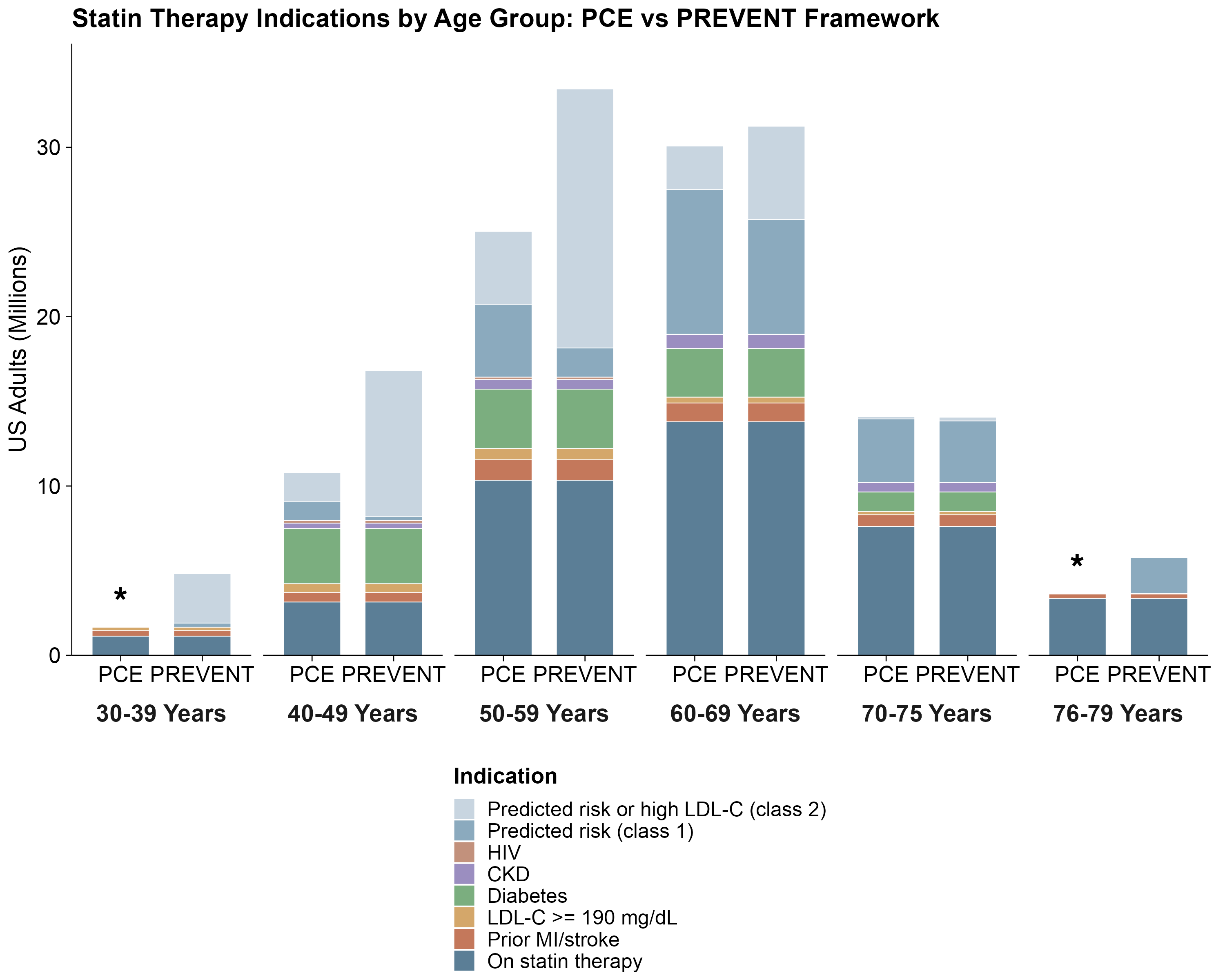}
\caption{\textbf{Statin eligibility by indication type and age group under 2018 and 2026 guidelines, US adults aged 30--79 years, NHANES 2011--2023.} Bars show the weighted number of US adults (millions) eligible by indication. ``Predicted risk (class~1)'' denotes eligibility at the class~1 threshold (PCE $\geq$7.5\% or PREVENT $\geq$5\%); ``Predicted risk or high LDL-C (class~2)'' collapses three class~2 indications: 10-year ASCVD risk at the class~2 threshold (PCE $\geq$5\% or PREVENT $\geq$3\%), LDL-C 160--189 mg/dL in adults aged 30--59 years, and 30-year PREVENT ASCVD risk $\geq$10\% in adults aged 30--59 years. For adults aged 40--75 years, risk-independent indications (statin use, ASCVD history, LDL-C $\geq$190 mg/dL, diabetes, CKD, HIV) are held constant across guidelines. Under both guidelines, adults aged 30--39 and 76--79 years were eligible through current statin use, LDL-C $\geq$190 mg/dL, or prior MI/stroke; under 2026 guidelines, these age groups also gain PREVENT-based risk eligibility. *Predicted risk based on PCE cannot be used to determine statin recommendations for these age groups because the PCEs are not validated for ages beyond 40--75 years. A corresponding figure with proportions on the y-axis is shown in Figure~\ref{fig:sfig_prop_barplot}.}
\label{fig:figure2}
\end{figure*}

Figure~\ref{fig:figure2} decomposes statin recommendations by indication type and age group under each guideline, with risk-independent indications were held constant across guidelines. Under 2026 guidelines, adults aged 30--39 and 76--79 years also became newly recommended for statin therapy through PREVENT-based risk thresholds. The reduction in risk--based recommendations among adults aged 40--75 years was most pronounced among those aged 50--69 years, where PCE assigns substantially higher absolute risk than PREVENT.

\subsection*{National Estimates for Statin Recommendations}

At the class~1 threshold, the number of US adults aged 40--75 years receiving or recommended for statin therapy decreased by an estimated 5.3 million (95\% CI, 4.8 million to 5.9 million), or $-$4.0 percentage points (pp), compared with 2018 guidelines. These reductions were partially offset by newly eligible adults in age groups where PREVENT extends beyond PCE: an estimated 0.2 million adults aged 30--39 years and 2.1 million adults aged 76--79 years became newly recommended for statin therapy. Across the full age range of 30--79 years, the net decrease was 3.0 million ($-$1.6pp; 95\% CI, $-$2.0 to $-$1.3). Decomposing this change, switching to PREVENT while retaining the 7.5\% threshold reduced statin recommendations by an estimated 10.4 million adults; the threshold reduction from 7.5\% to 5\% recaptured approximately 7.4 million adults, but did not fully offset the calculator effect.

When using the class~2 threshold, 86.2 million adults (47.8\%) were recommended for statin therapy under 2018 guidelines compared to 107.0 million (59.4\%) under 2026 guidelines, yielding a net increase of 20.8 million US adults (95\% CI, 19.6 million to 22.0 million) receiving or recommended for statins, or +11.6pp (95\% CI, 10.9 to 12.2). The class~2 threshold includes three new criteria for the 2026 framework beyond the 10-year risk threshold of PREVENT $\geq$3\%: LDL-C 160--189 mg/dL and 30-year PREVENT ASCVD risk $\geq$10\%, both applying to adults aged 30--59 years. The 30-year risk criterion was the dominant driver of this expansion, as 30-year PREVENT risk $\geq$10\% was met by the majority of adults aged 40--59 years. The increase was largest among adults aged 50--59 years (+19.7pp; 95\% CI, 18.0 to 21.4), followed by 40--49 years (+14.8pp; 95\% CI, 13.4 to 16.1), and 30--39 years (+7.8pp; 95\% CI, 6.7 to 8.9). Among adults aged 60--69 years, the class~2 threshold yielded a modest net gain (+3.3pp; 95\% CI, 2.3 to 4.3), as the 30-year risk criterion does not apply above age 59 years and effects are limited to the lower 10-year risk threshold.

Detreatment for lipid-lowering therapy (Figure~\ref{fig:sfig_detreat}), exclusion of participants currently receiving statin therapy (Figure~\ref{fig:sfig_nostatin}), restriction to participants with in-range equation inputs (Figure~\ref{fig:sfig_inrange}), and restriction to the fasting subsample (Figure~\ref{fig:sfig_fasting}) each produced net differences consistent with the primary analysis. Corresponding sensitivity analyses for the class~2 threshold yielded similarly consistent results (Figures~\ref{fig:sfig_detreat_c2}, \ref{fig:sfig_nostatin_c2}, \ref{fig:sfig_inrange_c2}, and \ref{fig:sfig_fasting_c2}).

\subsection*{Subgroup Differences}

\begin{figure*}[htbp]
\centering
\includegraphics[width=\textwidth]{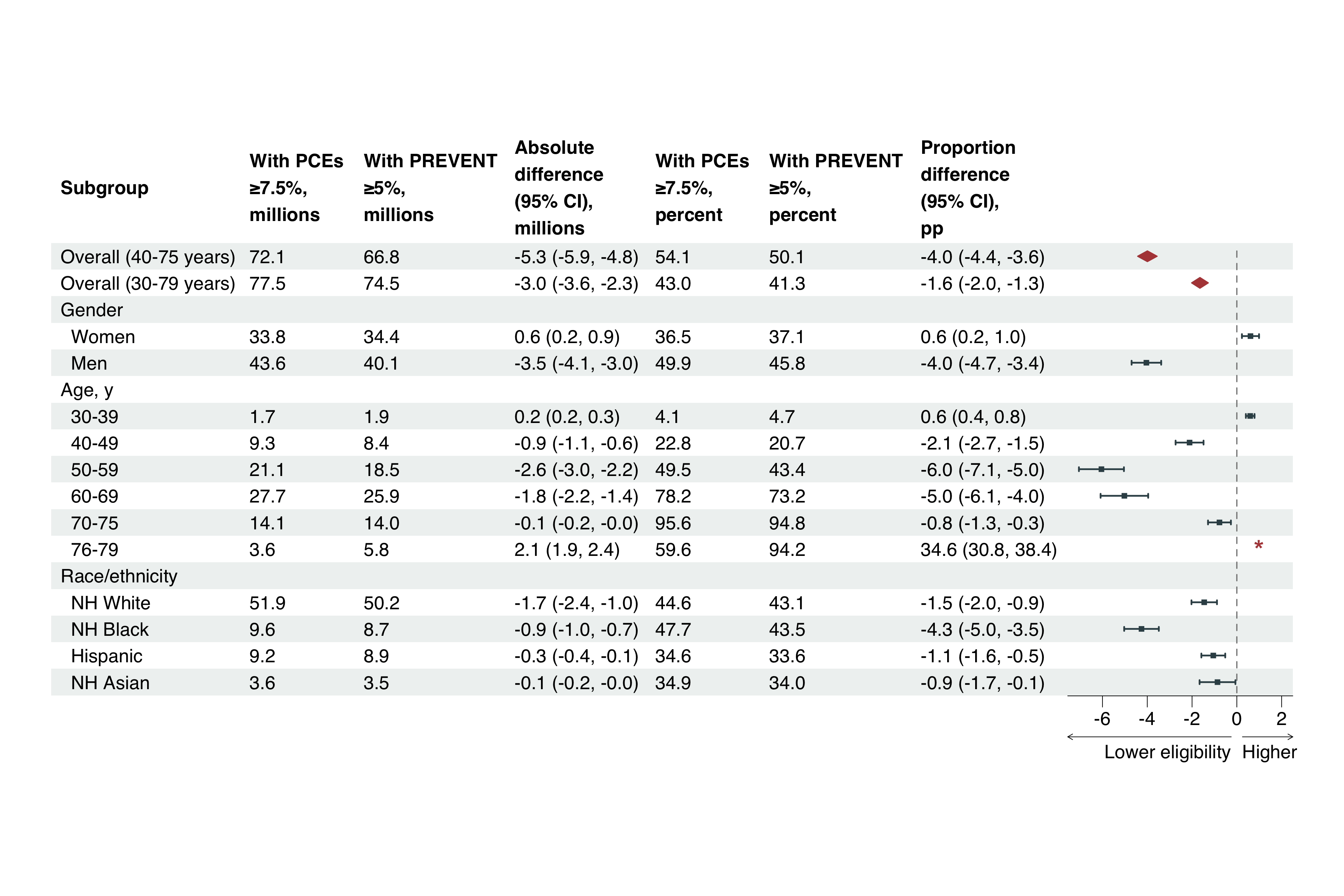}
\caption{\textbf{Net change in statin recommendations under 2026 guidelines vs 2018 guidelines: class~1 threshold (PREVENT $\geq$5\% vs PCE $\geq$7.5\%), US adults aged 30--79 years, NHANES 2011--2023.} Columns show absolute counts (millions) and proportions (\%) under each guideline; the rightmost column shows the proportion difference (percentage points) with 95\% CIs computed using paired within-person contrasts (see Methods). Negative values indicate decreased recommendations for statin therapy under 2026 guidelines. *The 76--79 age group is marked with an asterisk in the forest panel because the proportion difference falls far outside the range of other subgroups; this large effect reflects PREVENT extending risk-based recommendations to ages 76--79 years, whereas the PCEs are not validated beyond age 75.}
\label{fig:figure3}
\end{figure*}

\begin{figure*}[htbp]
\centering
\includegraphics[width=\textwidth]{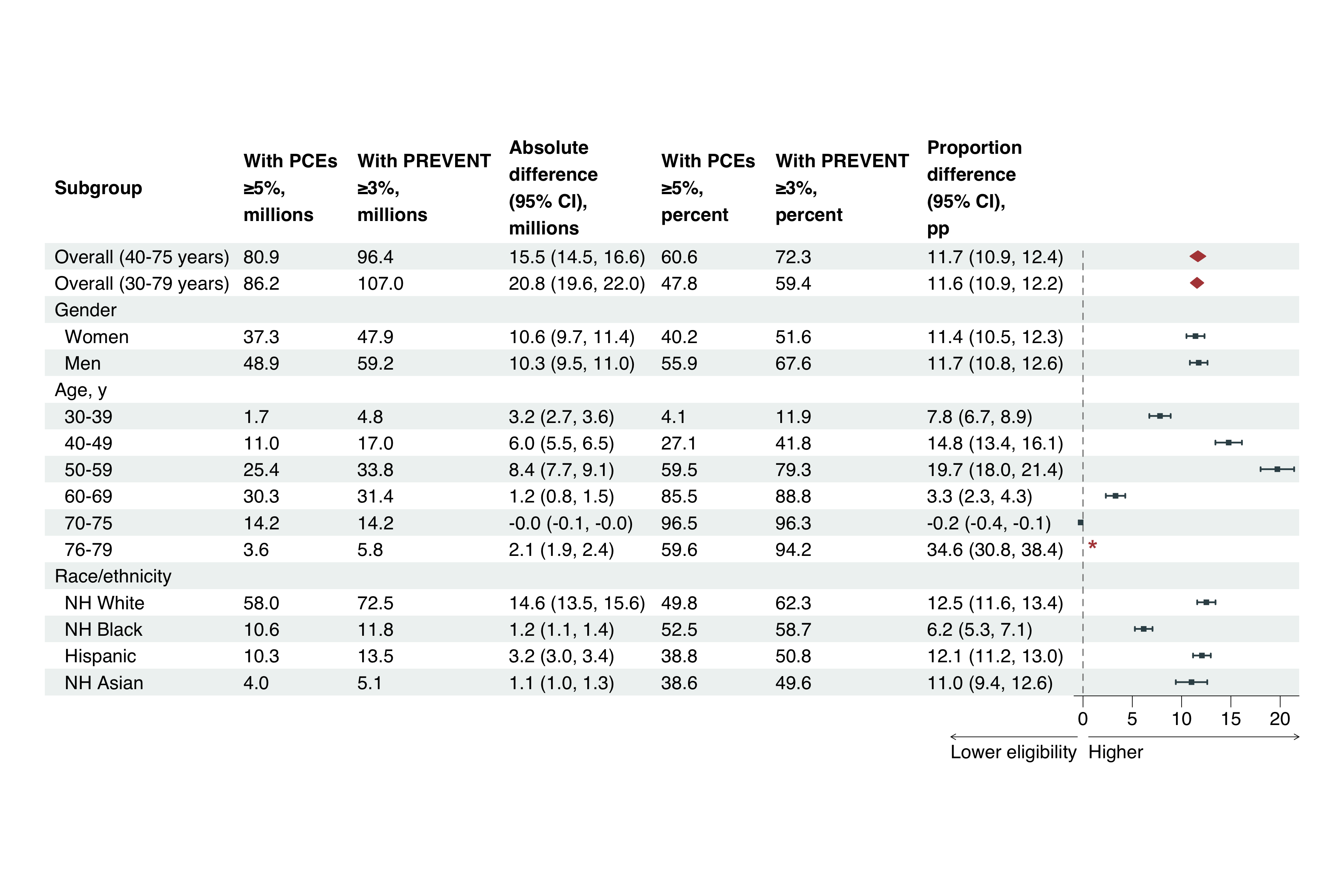}
\caption{\textbf{Net change in statin recommendations under 2026 guidelines vs 2018 guidelines: class~2 threshold (PREVENT $\geq$3\% + LDL-C 160--189 + 30-year ASCVD risk $\geq$10\% vs PCE $\geq$5\%), US adults aged 30--79 years, NHANES 2011--2023.} Under 2026 guidelines, the class~2 threshold additionally recommends statins for adults aged 30--59 years with LDL-C 160--189 mg/dL or 30-year PREVENT ASCVD risk $\geq$10\%. Columns show absolute counts (millions) and proportions (\%) under each guideline; the rightmost column shows the proportion difference (percentage points) with 95\% CIs computed using paired within-person contrasts (see Methods). Positive values indicate increased recommendations for statin therapy under 2026 guidelines. *The 76--79 age group is marked with an asterisk in the forest panel because the proportion difference falls far outside the range of other subgroups; this large effect reflects PREVENT extending risk-based recommendations to ages 76--79 years, whereas the PCEs are not validated beyond age 75.}
\label{fig:figure4}
\end{figure*}

The net effect of the 2026 guidelines at the class~1 threshold varied substantially across demographic subgroups (Figures~\ref{fig:figure3} and \ref{fig:figure4}). Men experienced a net decrease of 3.5 million ($-$4.0pp; 95\% CI, $-$4.7 to $-$3.4), whereas women experienced a net increase of 0.6 million (+0.6pp; 95\% CI, 0.2 to 1.0). Adults aged 50--59 years and 60--69 years experienced the largest proportional decreases ($-$6.0pp and $-$5.0pp). Adults aged 70--75 years experienced a smaller decrease ($-$0.8pp), as most were already recommended through risk-independent indications or high predicted risk. Adults aged 76--79 years experienced the largest proportional increase (+34.6pp, or 2.1 million; 95\% CI, 1.9 to 2.4 million). Adults aged 30--39 years showed a modest gain (+0.6pp, or 0.2 million) among the few who exceeded risk thresholds for PREVENT. Non-Hispanic Black adults experienced larger proportional reductions ($-$4.3pp; 95\% CI, $-$5.0 to $-$3.5). Non-Hispanic White, Hispanic, and Asian adults experienced smaller decreases of $-$1.5pp, $-$1.1pp, and $-$0.9pp.

At the class~2 threshold, the direction of the net effect reversed in most subgroups. Women gained an estimated 10.6 million (+11.4pp; 95\% CI, 10.5 to 12.3), and men gained 10.3 million (+11.7pp; 95\% CI, 10.8 to 12.6). Among racial/ethnic groups, non-Hispanic White adults gained the most in absolute terms (+14.6 million; +12.5pp), followed by Hispanic (+3.2 million; +12.1pp), NH Black (+1.2 million; +6.2pp), and NH Asian adults (+1.1 million; +11.0pp). Black adults experienced a smaller proportional increase at the class~2 threshold than other racial/ethnic groups, narrowing the disparity relative to the class~1 threshold but remaining a differential pattern warranting attention.

Subgroup analyses by poverty-income ratio, insurance status, educational attainment, and healthcare access showed consistent patterns across strata, with no evidence that the net changes in statin recommendations varied meaningfully by socioeconomic indicators (Figure~\ref{fig:sfig_ses}). In a multivariable survey-weighted linear regression among adults aged 40--75 years, Black race, older age, male gender, current smoking, and lower eGFR were independently more likely to no longer be recommended for statin therapy, while female gender and having a usual source of care were associated with a greater probability of being newly recommended for statin therapy (Figure~\ref{fig:sfig_regression}).

\section*{Discussion}

In this nationally representative analysis of US adults aged 30--79 years, we found that the simultaneous adoption of PREVENT equations and lower treatment thresholds introduced by the 2026 dyslipidemia guidelines produced a net reduction of 3.0 million statin-eligible adults ($-$1.6pp) with class~1 criteria and net increase of 20.8 million (+11.6pp) at the class~2 criteria, with substantial variation across difference age, gender, and race/ethnicity groups.

Our findings build on prior analyses \cite{Diao2024, Anderson2024} which estimated that substituting PREVENT for PCE at prior treatment thresholds would reduce statin recommendations by 14 million to 17 million adults. The 2026 guidelines partially offset this downward shift by lowering risk-based thresholds, but the class~1 threshold did not fully compensate: the net result was still a reduction of 3.0 million adults receiving or recommended for statin therapy. The class~2 threshold, by contrast, produced a net expansion of 20.8 million adults---driven not primarily by recalibration of 10-year risk categories but by the introduction of 30-year ASCVD risk assessment. The magnitude of this expansion---the largest since the adoption of risk-based guidelines in 2013---reflects a fundamental shift in the scope of primary prevention, from recalibrating existing risk categories to creating an entirely new pathway for statin consideration among middle-aged adults whose 10-year risk remains below conventional treatment thresholds. These changes follow prior guideline-driven shifts \cite{Stone2014, Pencina2014, Grundy2019} and represent the largest single expansion in statin eligibility since the 2013 risk-based framework was introduced.

Across all racial and ethnic groups, the direction of the net effect depended on the recommendation class. At the class~1 threshold, all groups experienced reductions in statin recommendations, but the magnitude differed substantially: Black adults experienced the largest proportional decrease ($-$4.3pp), more than double the overall effect. At the class~2 threshold, the direction reversed: all racial and ethnic groups gained statin recommendations, but Black adults experienced a smaller proportional increase---roughly half the gain seen in other groups. These disproportionate effects among Black adults are consistent with findings by Yan et al \cite{Yan2026} and likely reflect the removal of the PCE race coefficient, which assigned higher baseline ASCVD risk to Black adults \cite{Goff2014}. While removing race from risk calculators addresses an important algorithmic equity concern \cite{Vyas2020}, it also shifts the demographic distribution of recommended therapy in ways that warrant clinical vigilance. Black adults in the United States have persistently higher cardiovascular mortality rates than White adults \cite{Kyalwazi2022}, and guideline-concordant statin use for primary prevention is already lower among Black and Hispanic adults than among White adults across all risk strata \cite{Jacobs2023, Frank2023}. The disproportionate loss of formal recommendations at the class~1 threshold, combined with the smaller proportional gain at the class~2 threshold, suggests that gaps in recommended therapy for Black adults may widen under new guidelines. Clinicians may consider the use of shared decision-making tools, risk-enhancing factors, and coronary artery calcium scoring to ensure appropriate treatment recommendations where warranted \cite{Grundy2019}.

The guideline transition also produced divergent effects by sex. At the class~1 threshold, the net decrease was concentrated in men, while women experienced a slight net increase. This sex-based asymmetry may reflect the known tendency of PCE to overpredict risk in men compared to women, such that PREVENT's recalibration disproportionately lowers estimated risk---and thus statin eligibility---among men \cite{Rana2016, DeFilippis2017}. At the class~2 threshold, both sexes experienced substantial gains that were similar in magnitude. The convergence at the class~2 threshold reflects the 30-year risk criterion's broad applicability to both sexes, which effectively overwhelmed the sex-specific differences in 10-year risk recalibration affecting class~1 results. Statin therapy produces similar proportional reductions in major vascular events in women and men \cite{CTT2012}, yet women have historically been underrepresented in statin trials and in guideline-concordant statin prescribing. These changes therefore raise new questions about how the expanded recommendations may affect existing sex-based gaps in prescribing.

Age-related patterns further illustrate the divergent effects of the two recommendation classes. At the class~1 threshold, the largest proportional reductions occurred among middle-aged adults aged 50--69 years ($-$5.6pp), where PREVENT assigns substantially lower 10-year risk than PCE. Among adults aged 70--75 years, only a modest decrease was observed ($-$0.8pp), as most individuals were already recommended through risk-independent indications or high predicted risk under both frameworks. By contrast, adults aged 76--79 years experienced the largest proportional gain (+34.6pp; 2.1 million), because 2018 PCE-based guidelines did not assign risk-based recommendations beyond age 75 whereas 2026 guidelines apply PREVENT through age 79. Few adults aged 30--39 years became newly recommended based on 10-year risk predictions alone (+0.6pp).

The class~2 threshold reversed this age pattern. The 30-year risk criterion ($\geq$10\%) predominantly affected adults aged 40--59 years, who typically have low 10-year risk. Among adults aged 50--59 years, nearly 80\% would meet at least one statin criterion under the class~2 threshold, compared with 60\% under 2018 guidelines---an absolute increase of 19.7pp. Among adults aged 40--49 years, statin recommendations increased from 27\% to 42\%, adding an estimated 6.0 million newly recommended adults (+14.8pp). Even among adults aged 30--39 years, the class~2 criteria contributed a meaningful gain (+7.8pp). Among adults aged 60--69 years, the class~2 threshold yielded only a modest net gain (+3.3pp), as the 30-year risk criterion does not apply above age 59 years and effects are limited to the lower 10-year risk threshold. Adults aged 70--75 years showed essentially no net change ($-$0.2pp) at the class~2 threshold, as nearly all were already recommended under 2018 guidelines.

The implications of expanded class 1 recommendations among adults aged 76--79 years merit careful consideration. A meta-analysis of 28 randomized trials found that statin therapy significantly reduced major vascular events among older adults aged 65--75 years, but that this benefit was attenuated in those over 75 years \cite{CTT2019}. Adults aged 76--79 years are also more likely to have multimorbidity, polypharmacy, and frailty---factors that increase susceptibility to potential adverse effects \cite{NLA_AGS2024}. These considerations apply at both the class~1 and class~2 thresholds, though the absolute number of newly recommended older adults is substantially larger at the class~1 threshold because of PREVENT's extension of risk-based recommendations to this previously unscored age group.

This study has several limitations. First, using on-treatment lipid values may underestimate the net change in statin recommendations, because some current statin users would no longer meet PREVENT-based thresholds; our sensitivity analyses modeling cholesterol detreatment and with exclusion of current statin users yielded consistent results at both class~1 and class~2 thresholds. Second, guideline implementation involves clinical nuance---including shared decision-making, risk-enhancing factors, and coronary artery calcium scoring---that cannot be fully captured in a population-level analysis \cite{Grundy2019, Arnett2019}. This is particularly relevant for the class~2 threshold, where shared decision-making is expected to play a central role. Third, prescription drug data were not released in the 2021--2023 cycle, necessitating the use of self-reported cholesterol medication as a proxy for statin use in that cycle. Agreement between self-report and prescription data was high in earlier cycles (Supplementary Methods), but misclassification cannot be excluded. Fourth, LDL-C estimation required fasting triglycerides and was therefore available only in the fasting subsample; participants without LDL-C values were retained in the analysis but treated as not meeting LDL-C--based criteria rather than excluded. This approach may underestimate the prevalence of both severe hypercholesterolemia (LDL-C $\geq$190 mg/dL) and the class~2 LDL-C 160--189 mg/dL criterion. Our sensitivity analysis restricted to the fasting subsample with appropriate fasting weights showed that overall eligibility estimates were not substantially affected (Figure~\ref{fig:sfig_fasting}). Fifth, NHANES does not include validated 30-year ASCVD outcome data, so the accuracy of the PREVENT 30-year risk equation in this population cannot be assessed; the substantial proportion of middle-aged adults meeting the $\geq$10\% threshold underscores the importance of external validation studies. Finally, our analysis does not account for the recent stalling of cardiovascular mortality declines in the US, which may alter the population risk distribution and the absolute benefit of preventive therapy over time.

\section*{Conclusions}

The population impact of the 2026 dyslipidemia guidelines depends critically on which recommendation class is applied. At the class~1 threshold, adoption of PREVENT with lower treatment thresholds may reduce the number of US adults receiving or recommended for statin therapy by an estimated 3.0 million ($-$1.6pp), with larger proportional decreases among Black, male, and middle-aged adults. At the class~2 threshold, the consideration of 30-year ASCVD risk assessment may increase statin recommendations by an estimated 20.8 million (+11.6pp), representing the largest single expansion in statin eligibility since the introduction of risk-based guidelines. The choice between class 1 and class 2 thresholds is not yet standardized across health systems, and variation in institutional adoption could itself become a driver of care disparities, making transparent guideline implementation tracking a priority. Our findings underscore the need for monitoring of prescribing, outcomes, and equity during the guideline transition. 

\bibliographystyle{unsrt}
\bibliography{references}

\clearpage


\begin{table*}[!t]
\centering
\caption{\textbf{Weighted Characteristics of US Adults Aged 30--79 Years, Overall and by Statin Eligibility Under 2018 and 2026 Guidelines at Class~1 and Class~2 Thresholds, NHANES 2011--2023.} ``Eligible 2018'' = receiving or recommended for statin therapy under 2018 PCE-based guidelines. ``Eligible 2026'' = receiving or recommended under 2026 PREVENT-based guidelines. Class~1 threshold: PCE $\geq$7.5\% vs PREVENT $\geq$5\%. Class~2 threshold: PCE $\geq$5\% vs PREVENT $\geq$3\%, plus LDL-C 160--189 mg/dL or 30-year PREVENT ASCVD risk $\geq$10\% in adults aged 30--59 years. Continuous variables are reported as median (IQR); categorical variables as weighted \% (SE). The ``Difference'' columns show the difference between 2026-eligible and 2018-eligible domain means (mean difference for continuous variables; proportion difference in percentage points for categorical variables) with 95\% CIs accounting for within-person covariance between the overlapping subpopulations (see Methods).}
\label{tab:table1}
\scriptsize
\setlength{\tabcolsep}{3pt}
\renewcommand{\arraystretch}{0.95}
\begin{tabularx}{\textwidth}{@{}>{\raggedright\arraybackslash}p{3.0cm} *{7}{Y}@{}}
\toprule
 & & \multicolumn{3}{c}{Class 1 (7.5\%/5\%)} & \multicolumn{3}{c}{Class 2 (5\%/3\%+)} \\
\cmidrule(lr){3-5} \cmidrule(lr){6-8}
Variable & Overall & \shortstack{Eligible\\2018} & \shortstack{Eligible\\2026} & \shortstack{Difference\\(95\% CI)} & \shortstack{Eligible\\2018} & \shortstack{Eligible\\2026} & \shortstack{Difference\\(95\% CI)} \\
\midrule
N (unweighted) & 24{,}199 & 11{,}806 & 11{,}423 & $-$383 & 12{,}891 & 15{,}256 & 2{,}365 \\
N (weighted, millions) & 180.2 & 77.5 & 74.5 & $-$3.0 & 86.2 & 107.0 & 20.8 \\
\addlinespace
Age, median (IQR), y & 52.0 (41.0--62.0) & 62.0 (55.0--69.0) & 63.0 (55.0--70.0) & 0.7 (0.6, 0.8) & 61.0 (54.0--68.0) & 59.0 (51.0--67.0) & $-$1.7 ($-$1.9, $-$1.6) \\
\addlinespace
\multicolumn{8}{l}{\textit{Age group, \%}} \\
\quad 30--39 y & 22.5 (0.5) & 2.2 (0.2) & 2.6 (0.2) & 0.4 (0.3, 0.5) & 1.9 (0.2) & 4.5 (0.3) & 2.6 (2.1, 3.0) \\
\quad 40--49 y & 22.6 (0.4) & 12.0 (0.4) & 11.3 (0.4) & $-$0.7 ($-$1.0, $-$0.4) & 12.8 (0.4) & 15.9 (0.4) & 3.1 (2.6, 3.7) \\
\quad 50--59 y & 23.7 (0.4) & 27.3 (0.7) & 24.9 (0.7) & $-$2.4 ($-$2.9, $-$1.9) & 29.5 (0.7) & 31.6 (0.6) & 2.1 (1.5, 2.7) \\
\quad 60--69 y & 19.7 (0.5) & 35.8 (0.7) & 34.8 (0.7) & $-$1.0 ($-$1.4, $-$0.5) & 35.1 (0.7) & 29.4 (0.6) & $-$5.7 ($-$6.2, $-$5.3) \\
\quad 70--75 y & 8.2 (0.3) & 18.2 (0.5) & 18.7 (0.6) & 0.6 (0.4, 0.8) & 16.5 (0.5) & 13.2 (0.4) & $-$3.2 ($-$3.5, $-$3.0) \\
\quad 76--79 y & 3.4 (0.1) & 4.7 (0.3) & 7.7 (0.3) & 3.0 (2.7, 3.4) & 4.2 (0.2) & 5.4 (0.2) & 1.2 (0.9, 1.4) \\
\addlinespace
Female, \% & 51.5 (0.4) & 43.7 (0.6) & 46.2 (0.6) & 2.5 (2.1, 2.9) & 43.3 (0.6) & 44.7 (0.5) & 1.5 (0.9, 2.1) \\
\addlinespace
\multicolumn{8}{l}{\textit{Race/ethnicity}} \\
\quad Hispanic & 14.7 (1.0) & 11.9 (1.0) & 11.9 (1.0) & 0.1 ($-$0.1, 0.3) & 11.9 (1.0) & 12.6 (1.0) & 0.7 (0.4, 0.9) \\
\quad NH Asian & 5.7 (0.5) & 4.7 (0.4) & 4.7 (0.4) & 0.1 ($-$0.0, 0.2) & 4.6 (0.4) & 4.8 (0.4) & 0.2 ($-$0.0, 0.3) \\
\quad NH Black & 11.2 (0.8) & 12.4 (0.9) & 11.7 (0.9) & $-$0.7 ($-$0.9, $-$0.4) & 12.2 (0.9) & 11.0 (0.8) & $-$1.2 ($-$1.5, $-$1.0) \\
\quad NH White & 64.6 (1.4) & 67.0 (1.4) & 67.4 (1.5) & 0.4 ($-$0.0, 0.8) & 67.2 (1.4) & 67.7 (1.4) & 0.5 (0.0, 1.0) \\
\quad Other & 3.8 (0.2) & 4.1 (0.3) & 4.2 (0.4) & 0.1 ($-$0.1, 0.3) & 4.0 (0.3) & 3.9 (0.3) & $-$0.1 ($-$0.3, 0.1) \\
\addlinespace
BMI, median (IQR), kg/m$^2$ & 28.6 (25.0--33.4) & 29.6 (26.0--34.3) & 29.7 (26.1--34.5) & 0.1 (0.1, 0.2) & 29.4 (25.9--34.1) & 29.4 (25.9--34.1) & 0.0 ($-$0.1, 0.1) \\
SBP, median (IQR), mmHg & 121.0 (111.5--132.0) & 127.0 (117.0--139.0) & 127.0 (117.0--139.5) & 0.3 (0.1, 0.4) & 126.7 (116.5--138.0) & 126.0 (116.0--137.5) & $-$0.5 ($-$0.7, $-$0.4) \\
Total cholesterol, mg/dL & 192.0 (166.0--219.0) & 188.0 (160.0--219.0) & 187.0 (158.0--218.0) & $-$0.8 ($-$1.2, $-$0.4) & 191.0 (162.0--221.0) & 196.0 (166.0--226.0) & 4.0 (3.5, 4.5) \\
HDL-C, median (IQR), mg/dL & 51.0 (42.0--63.0) & 49.0 (41.0--60.0) & 49.0 (41.0--60.0) & $-$0.0 ($-$0.1, 0.1) & 49.0 (41.0--60.0) & 49.0 (41.0--60.0) & $-$0.3 ($-$0.4, $-$0.1) \\
LDL-C, median (IQR), mg/dL & 114.0 (91.0--138.0) & 109.7 (85.0--137.0) & 109.0 (84.4--137.0) & $-$0.5 ($-$0.9, $-$0.1) & 112.1 (87.0--139.0) & 117.0 (92.0--146.0) & 4.2 (3.5, 4.9) \\
Triglycerides, mg/dL & 101.0 (70.0--149.0) & 114.0 (80.0--166.0) & 115.0 (82.0--167.0) & $-$0.4 ($-$2.0, 1.1) & 113.0 (79.0--165.0) & 113.0 (80.0--165.0) & $-$0.3 ($-$2.0, 1.5) \\
eGFR, mL/min/1.73m$^2$ & 95.7 (81.4--107.0) & 88.1 (72.4--99.3) & 87.0 (71.2--98.7) & $-$1.0 ($-$1.2, $-$0.8) & 88.8 (73.5--99.6) & 90.2 (75.1--101.0) & 1.5 (1.3, 1.7) \\
HbA1c, median (IQR), \% & 5.5 (5.3--5.9) & 5.8 (5.5--6.4) & 5.8 (5.5--6.5) & 0.0 (0.0, 0.0) & 5.7 (5.4--6.3) & 5.7 (5.4--6.1) & $-$0.1 ($-$0.1, $-$0.1) \\
\addlinespace
Current smoker, \% & 18.4 (0.5) & 20.4 (0.6) & 18.6 (0.6) & $-$1.8 ($-$2.2, $-$1.4) & 21.2 (0.6) & 19.9 (0.6) & $-$1.3 ($-$1.8, $-$0.8) \\
Diabetes, \% & 16.9 (0.4) & 36.5 (0.7) & 38.6 (0.7) & 2.2 (1.8, 2.5) & 32.8 (0.6) & 27.6 (0.6) & $-$5.2 ($-$5.7, $-$4.8) \\
CKD stage 3--4, \% & 5.0 (0.2) & 10.6 (0.4) & 11.8 (0.4) & 1.2 (1.0, 1.4) & 9.5 (0.4) & 8.2 (0.3) & $-$1.3 ($-$1.5, $-$1.1) \\
HIV, \% & 0.3 (0.1) & 0.6 (0.1) & 0.6 (0.1) & 0.0 (0.0, 0.0) & 0.6 (0.1) & 0.4 (0.1) & $-$0.1 ($-$0.2, $-$0.1) \\
ASCVD history, \% & 8.9 (0.3) & 20.2 (0.6) & 21.0 (0.6) & 0.8 (0.6, 1.0) & 18.2 (0.5) & 14.7 (0.4) & $-$3.5 ($-$3.8, $-$3.3) \\
BP medication use, \% & 32.4 (0.5) & 57.3 (0.7) & 59.7 (0.7) & 2.3 (1.9, 2.8) & 54.1 (0.7) & 48.3 (0.6) & $-$5.7 ($-$6.4, $-$5.1) \\
Statin use, \% & 21.9 (0.5) & 50.9 (0.8) & 52.9 (0.8) & 2.0 (1.6, 2.5) & 45.7 (0.7) & 36.8 (0.7) & $-$8.9 ($-$9.4, $-$8.4) \\
LDL-C $\geq$190 mg/dL, \% & 1.3 (0.1) & 3.0 (0.3) & 3.1 (0.3) & 0.1 (0.1, 0.2) & 2.7 (0.2) & 2.2 (0.2) & $-$0.5 ($-$0.6, $-$0.4) \\
\addlinespace
10-y ASCVD risk (PCE), \% & 6.0 (2.2--13.9) & 11.8 (7.4--19.7) & 13.0 (6.9--21.4) & 0.9 (0.9, 1.0) & 10.6 (6.3--18.4) & 8.8 (4.6--17.1) & $-$1.4 ($-$1.5, $-$1.2) \\
10-y ASCVD risk (PREVENT), \% & 2.5 (0.9--5.9) & 6.2 (4.0--9.3) & 6.7 (4.2--10.0) & 0.4 (0.4, 0.5) & 5.6 (3.6--8.9) & 4.7 (2.7--8.2) & $-$0.7 ($-$0.7, $-$0.6) \\
\bottomrule
\end{tabularx}
\end{table*}

\FloatBarrier

\clearpage
\appendix
\section*{Supplementary Appendix}

\subsection*{Supplementary Methods}

\subsubsection*{Data Sources and Harmonization}

We pooled data from five consecutive NHANES cycles: 2011--2012, 2013--2014, 2015--2016, 2017--March 2020 (prepandemic), and 2021--2023. For each cycle, we extracted the following datasets: demographics (DEMO), blood pressure examination (BPX/BPXO), body measures (BMX), serum triglycerides (TRIGLY), total cholesterol (TCHOL), HDL cholesterol (HDL), biochemistry profile (BIOPRO, for serum creatinine), glycohemoglobin (GHB, for HbA1c), fasting plasma glucose (GLU), albumin and creatinine in urine (ALB\_CR), diabetes questionnaire (DIQ), blood pressure and cholesterol questionnaire (BPQ), smoking questionnaire (SMQ), medical conditions questionnaire (MCQ), prescription medications (RXQ\_RX), kidney conditions (KIQ\_U), HIV laboratory results (HIV), and health insurance (HIQ). Data files were merged by NHANES sequence number (SEQN) within each cycle; all cycles were then combined into a single analytic dataset. Variables not present in a given cycle (e.g., RXDDRUG in 2021--2023) were set to missing and handled through cycle-specific logic as described below.

\subsubsection*{Variable Construction}

\textbf{Blood pressure.} Systolic and diastolic blood pressure were computed as the mean of available readings from the second through fourth measurements. In cycles using oscillometric devices (2017--March 2020, 2021--2023), readings from BPXOSY2, BPXOSY3 were used; in earlier cycles, readings from BPXSY2, BPXSY3, and BPXSY4 were used. Diastolic blood pressure values of exactly 0 were recoded as missing. Where oscillometric and auscultatory variables coexisted, oscillometric values were prioritized using \texttt{coalesce()}.

\textbf{LDL cholesterol estimation.} LDL-C was estimated using the Martin--Hopkins method for all participants with available total cholesterol, HDL-C, and triglycerides, regardless of fasting status. The Martin--Hopkins method uses a 180-cell (30 triglyceride strata $\times$ 6 non-HDL cholesterol strata) adjustable factor table to compute the VLDL-C/triglyceride ratio, from which LDL-C is calculated as total cholesterol $-$ HDL-C $-$ triglycerides/adjustment factor. Triglyceride strata ranged from $<$50 to $\geq$400 mg/dL, and non-HDL cholesterol strata ranged from $<$100 to $\geq$220 mg/dL, each mapped to a specific divisor from the published 30$\times$6 factor matrix. Where NHANES-provided Martin--Hopkins LDL-C (LBDLDLM) was available, it was used directly; otherwise, LDL-C was computed from the factor table using measured total cholesterol, HDL-C, and triglycerides.

\textbf{eGFR.} Estimated glomerular filtration rate was calculated using the CKD-EPI 2021 race-free creatinine equation, implemented via the \texttt{kidney.epi} R package (\texttt{egfr.ckdepi.cr.2021}), using serum creatinine (LBXSCR) and demographic inputs (age, sex). CKD stage 3--4 was defined as eGFR 15--59 mL/min/1.73m$^2$.

\textbf{Diabetes.} Diabetes was defined as any of: (1) self-reported physician diagnosis (DIQ010 = ``Yes''), (2) use of a glucose-lowering medication identified from RXDDRUG, (3) HbA1c $\geq$6.5\%, or (4) fasting glucose $\geq$126 mg/dL. The glucose-lowering medication list included metformin, sulfonylureas (glipizide, glyburide, glimepiride), thiazolidinediones, dipeptidyl peptidase-4 (DPP-4) inhibitors, sodium-glucose cotransporter-2 (SGLT2) inhibitors, glucagon-like peptide-1 (GLP-1) receptor agonists, insulin preparations, and other agents. A narrower definition restricted to diagnosed diabetes (self-report or medication use) was also retained.

\textbf{Smoking.} Current smoking was defined as a ``Yes'' response to SMQ020 (smoked $\geq$100 cigarettes in lifetime) combined with a response of ``Every day'' or ``Some days'' to SMQ040 (currently smoking frequency).

\textbf{Clinical ASCVD history.} ASCVD history was defined as a ``Yes'' response to any of the medical conditions questionnaire items for coronary heart disease (MCQ160C), angina (MCQ160D), myocardial infarction (MCQ160E), or stroke (MCQ160F).

\textbf{HIV status.} HIV status was ascertained from laboratory results (LBDHI = ``Positive'', LBXHIVC = ``Reactive'', or LBXHIV1 = ``Reactive'') or antiretroviral medication use identified from RXDDRUG. The antiretroviral list included nucleoside/nucleotide reverse transcriptase inhibitors, non-nucleoside reverse transcriptase inhibitors, protease inhibitors, integrase strand transfer inhibitors, entry inhibitors, pharmacokinetic enhancers, and fixed-dose combination products (e.g., Biktarvy, Triumeq, Genvoya, Symtuza).

\textbf{Severe hypercholesterolemia.} Defined as Martin--Hopkins LDL-C $\geq$190 mg/dL, ascertained among all participants with available total cholesterol, HDL-C, and triglycerides regardless of fasting status.

\subsubsection*{Medication Identification}

In NHANES cycles 2011--2012 through 2017--March 2020, all prescription medications used in the past 30 days were recorded with generic drug names (RXDDRUG). Statins were identified by matching generic names for atorvastatin, rosuvastatin, simvastatin, pravastatin, lovastatin, fluvastatin, pitavastatin, and cerivastatin. Other lipid-lowering agents were classified into ezetimibe, PCSK9 inhibitors (alirocumab, evolocumab, inclisiran), fibrates (fenofibrate, gemfibrozil), niacin, bile acid sequestrants (cholestyramine, colesevelam, colestipol), and bempedoic acid. Antihypertensive medications were identified across multiple drug classes including angiotensin-converting enzyme inhibitors, angiotensin receptor blockers, calcium channel blockers, thiazide and loop diuretics, potassium-sparing diuretics, beta-blockers, alpha-blockers, centrally-acting agents, and direct vasodilators.

For the 2021--2023 cycle, prescription drug generic names were not released. Statin use was therefore ascertained from self-reported cholesterol medication use (BPQ100D or BPQ101D = ``Yes''). In earlier cycles where both data sources were available, self-reported cholesterol medication showed 94.9\% agreement with prescription-identified statin use (Cohen's $\kappa$ = 0.85), with sensitivity of 0.893 and specificity of 0.965 (see below). Lipid-lowering medication use for the detreatment sensitivity analysis was similarly proxied by self-reported cholesterol medication in this cycle.

\subsubsection*{Risk Score Computation}

\textbf{Pooled Cohort Equations (PCE).} The 10-year ASCVD risk was computed using the sex- and race-specific Goff 2013 equations via the \texttt{PooledCohort} R package (\texttt{predict\_10yr\_ascvd\_risk} with \texttt{equation\_version = ``Goff\_2013''}) and verified against the American College of Cardiology (ACC) Risk Estimator. Required inputs include age, sex, race, total cholesterol, HDL-C, systolic blood pressure, blood pressure treatment, smoking, and diabetes. Continuous inputs were capped at the recommended ranges: total cholesterol 130--320 mg/dL, HDL-C 20--100 mg/dL, systolic blood pressure 90--200 mmHg, and age 40--79 years. Although the PCEs were developed with non-Hispanic Black and non-Hispanic White as the only two race options, we followed guideline recommendations \cite{Goff2014} and used the White coefficients for Hispanic, Asian, and other-race participants, consistent with commonly used online implementations. Adults aged 30--39 years were not assigned PCE scores because these equations are not validated below age 40.

\textbf{PREVENT equations.} The 10-year PREVENT-ASCVD risk was computed using the Khan 2023 base model via the \texttt{PooledCohort} R package (\texttt{predict\_10yr\_ascvd\_risk} with \texttt{equation\_version = ``Khan\_2023''} and \texttt{prevent\_type = ``base''}) and verified against the AHA PREVENT online calculator. The PREVENT equations share several inputs with PCE but additionally require eGFR, BMI, and statin use, and remove race as an input. PREVENT also accepts HbA1c, urine albumin-to-creatinine ratio (UACR), and social deprivation index as optional inputs, but we used only the base equations without these optional inputs. PREVENT was computed for all adults aged 30--79 years with complete data on age, sex, total cholesterol, HDL-C, systolic blood pressure, blood pressure treatment, statin use, diabetes, smoking, BMI, and eGFR. Continuous inputs were capped: total cholesterol 130--320 mg/dL, HDL-C 20--100 mg/dL, systolic blood pressure 90--200 mmHg, BMI 15--60 kg/m$^2$, eGFR 15--140 mL/min/1.73m$^2$, and age 30--79 years. To reflect common clinical practice, we used the nearest acceptable value when input values fell outside these ranges rather than excluding observations \cite{Diao2024, Anderson2024}.

\textbf{PREVENT 30-year ASCVD risk.} The 30-year PREVENT-ASCVD risk was computed using the Khan 2023 base model via the \texttt{PooledCohort} R package (\texttt{predict\_30yr\_ascvd\_risk} with \texttt{equation\_version = ``Khan\_2023''} and \texttt{prevent\_type = ``base''}). The 30-year risk equation uses the same inputs as the 10-year PREVENT model. This 30-year risk estimate was used for the class~2 recommendation criterion: adults aged 30--59 years with 30-year PREVENT ASCVD risk $\geq$10\% are recommended for statin consideration under the 2026 dyslipidemia guidelines.

\subsubsection*{Decomposition by Reason for Statin Eligibility}

For the stacked barplot visualizations (Figure~2 and Figure~\ref{fig:sfig_prop_barplot}), each participant's primary reason for statin eligibility was assigned using a priority ordering: (1) on statin therapy, (2) prior myocardial infarction (MI) or stroke, (3) LDL-C $\geq$190 mg/dL, (4) diabetes (ages 40--75), (5) CKD stage 3--4 (ages 40--75), (6) HIV (ages 40--75), (7) predicted 10-year risk at class~1 threshold, (8) class~2 indications (LDL-C 160--189 mg/dL in ages 30--59, 30-year ASCVD risk $\geq$10\% in ages 30--59, or predicted 10-year risk at class~2 threshold). For display purposes, the three class~2 sub-categories were collapsed into a single ``Predicted risk or high LDL-C (class~2)'' category. Participants not meeting any criterion were classified as ``Not eligible.'' This decomposition was performed separately under the PCE framework and the PREVENT framework.

\subsubsection*{Categorization of Race and Ethnicity}

Categories for race and ethnicity were mutually exclusive and derived from self-reported survey responses (RIDRETH3 when available, otherwise RIDRETH1). Respondents who self-identified as ``Mexican American'' or ``Other Hispanic'' were categorized as Hispanic. The remaining participants were categorized as non-Hispanic White, non-Hispanic Black, non-Hispanic Asian, or Other Race (including multi-racial). Participants reporting ``Other Race -- Including Multi-Racial'' were included in overall analyses but not in race/ethnicity-specific subgroup comparisons.

\subsubsection*{Survey Weight Construction and Missing Data}

All analyses used NHANES MEC examination weights. Because the five pooled cycles spanned unequal time intervals (four 2-year cycles and one 3.2-year prepandemic cycle), survey weights were adjusted following National Center for Health Statistics guidance for pooling cycles of unequal length. Each cycle's MEC weight was multiplied by the ratio of its duration (in years) to the total pooled time span: the 2017--March 2020 prepandemic file was assigned 3.2 years, and all other cycles were assigned 2 years, for a total of 11.2 years. The pooled weight was therefore $w_{\text{pooled}} = w_{\text{MEC}} \times (\text{cycle years} / 11.2)$. Fasting-subsample weights were similarly pooled for the fasting sensitivity analysis. Variance estimation used Taylor series linearization with stratification (SDMVSTRA) and clustering (SDMVPSU) variables, implemented via the \texttt{survey} R package with \texttt{svydesign(ids = \textasciitilde SDMVPSU, strata = \textasciitilde SDMVSTRA, weights = \textasciitilde wt\_mec\_pooled, nest = TRUE)}. Confidence intervals were computed using linearization-based variance estimation. Because key analytic variables were missing for fewer than 10\% of participants, observations with missing values for risk score inputs were excluded from risk-based eligibility classification without additional re-weighting or imputation, consistent with NHANES analytic guidelines \cite{Akinbami2022}. One notable exception is LDL-C, which requires fasting triglycerides for Martin--Hopkins estimation and was therefore available only in the fasting subsample (approximately 47\% of the analytic cohort). Participants without LDL-C measurements were retained in the analysis but treated as not meeting LDL-C--based criteria (LDL-C $\geq$190 mg/dL and LDL-C 160--189 mg/dL); this approach may underestimate the prevalence of these indications in the full MEC sample.

\subsubsection*{Detreatment Sensitivity Analysis}

To approximate untreated baseline cholesterol values among participants receiving lipid-lowering therapy, we applied multiplicative correction factors reflecting average class-specific treatment effects. Observed values were modeled as: $\text{observed} = \text{untreated} \times \prod_i (1 - r_i)$, where $r_i$ is the proportional reduction attributable to each medication class. The following average reductions were assumed: statins (LDL-C $-$35\%, total cholesterol $-$25\%), ezetimibe (LDL-C $-$18\%, total cholesterol $-$13\%), PCSK9 inhibitors (LDL-C $-$60\%, total cholesterol $-$36\%), fibrates (LDL-C $-$8\%, total cholesterol $-$10\%, triglycerides $-$30\%), and niacin (LDL-C $-$15\%, total cholesterol $-$10\%, triglycerides $-$20\%). For participants on combination therapy, effects were applied multiplicatively. Untreated values were recovered as $\text{untreated} = \text{observed} / \prod_i (1 - r_i)$. HDL-C was not adjusted. After adjustment, risk scores were recomputed using the adjusted lipid values, and LDL-C $\geq$190 mg/dL was re-evaluated. For the PREVENT recomputation, the statin use indicator was set to ``no'' to estimate risk in the hypothetical untreated state. For the 2021--2023 cycle, in which prescription drug generic names were not released, individual non-statin lipid-lowering agents (ezetimibe, PCSK9 inhibitors, fibrates, niacin) could not be identified; the detreatment adjustment for these participants therefore applied only the statin correction factor, based on self-reported cholesterol medication use, and did not account for potential concomitant non-statin therapy. This detreatment analysis is limited to fixed, class-average multiplicative adjustment factors because NHANES does not capture statin intensity (dose), adherence, or duration of therapy; however, at the population level we expect these average corrections to provide an unbiased estimate of the direction and approximate magnitude of baseline cholesterol levels.

\subsubsection*{In-Range Equation Input Sensitivity Analysis}

For the in-range sensitivity analysis, we excluded participants with any risk factor value outside the recommended input ranges for either calculator. Out-of-range criteria for PCE included total cholesterol $<$130 or $>$320 mg/dL, HDL-C $<$20 or $>$100 mg/dL, systolic blood pressure $<$90 or $>$200 mmHg, or age $<$40 or $>$79 years. Out-of-range criteria for PREVENT additionally included BMI $<$15 or $>$60 kg/m$^2$ and eGFR $<$15 or $>$140 mL/min/1.73m$^2$. Participants were excluded if they met out-of-range criteria for \emph{either} calculator.

\subsubsection*{Socioeconomic Subgroup Definitions}

Poverty-income ratio (PIR) was categorized as $<$1.0 (below federal poverty level), 1.0--1.99, 2.0--3.99, and $\geq$4.0, using the NHANES family poverty-income ratio variable (INDFMPIR). Insurance status was ascertained from HIQ011 (``covered by health insurance''). Education was categorized as less than high school (less than 9th grade, or 9th--11th grade without diploma), high school graduate/GED, some college or associate degree, and college graduate or above, based on DMDEDUC2. NHANES cycle subgroups used the five pooled cycles as defined above.

\subsubsection*{Agreement Between Self-Reported Cholesterol Medication and Prescription-Identified Statin Use}

Among 19{,}153 participants in NHANES cycles 2011--2012 through 2017--March 2020 with non-missing values for both self-reported cholesterol medication use (BPQ100D/BPQ101D) and prescription-identified statin use (RXDDRUG), overall discordance was 5.1\% (980 of 19{,}153). The overall agreement was 94.9\%, with sensitivity of 0.893, specificity of 0.965, and Cohen's $\kappa$ of 0.85. Discordance was consistently below 5.5\% across all cycles: 5.2\% in 2011--2012 (n = 4{,}013), 4.6\% in 2013--2014 (n = 4{,}315), 4.9\% in 2015--2016 (n = 4{,}169), and 5.5\% in 2017--March 2020 (n = 6{,}656). These results support the use of self-reported cholesterol medication as a proxy for statin use in the 2021--2023 cycle, in which prescription drug names were not released.

\subsubsection*{Software and Reproducibility}

All analyses were conducted in R version 4.4. Key R packages included \texttt{survey} (version 4.4-2) for complex survey estimation, \texttt{PooledCohort} for PCE and PREVENT risk score computation, \texttt{kidney.epi} for CKD-EPI 2021 eGFR calculation, \texttt{forestploter} for forest plot visualization, \texttt{ggplot2} and \texttt{ggalluvial} for heatmap and alluvial figures, and \texttt{patchwork} for figure composition.

\clearpage
\setcounter{figure}{0}
\renewcommand{\thefigure}{S\arabic{figure}}
\subsection*{Supplementary Figures}

\begin{figure*}[htbp]
\centering
\caption{\textbf{Subgroup forest plot: net change in statin recommendations by socioeconomic and NHANES cycle subgroups, class~1 threshold (PREVENT $\geq$5\% vs PCE $\geq$7.5\%), US adults aged 30--79 years, NHANES 2011--2023.} Subgroups include poverty-income ratio categories, insurance status, educational attainment, country of birth (US-born vs foreign-born), language spoken at home (English only vs non-English or bilingual), usual source of health care, doctor visit in the past 12 months, and NHANES cycle. The 95\% CIs use the same paired within-person contrast approach as Figure~\ref{fig:figure3}.}
\includegraphics[width=\textwidth]{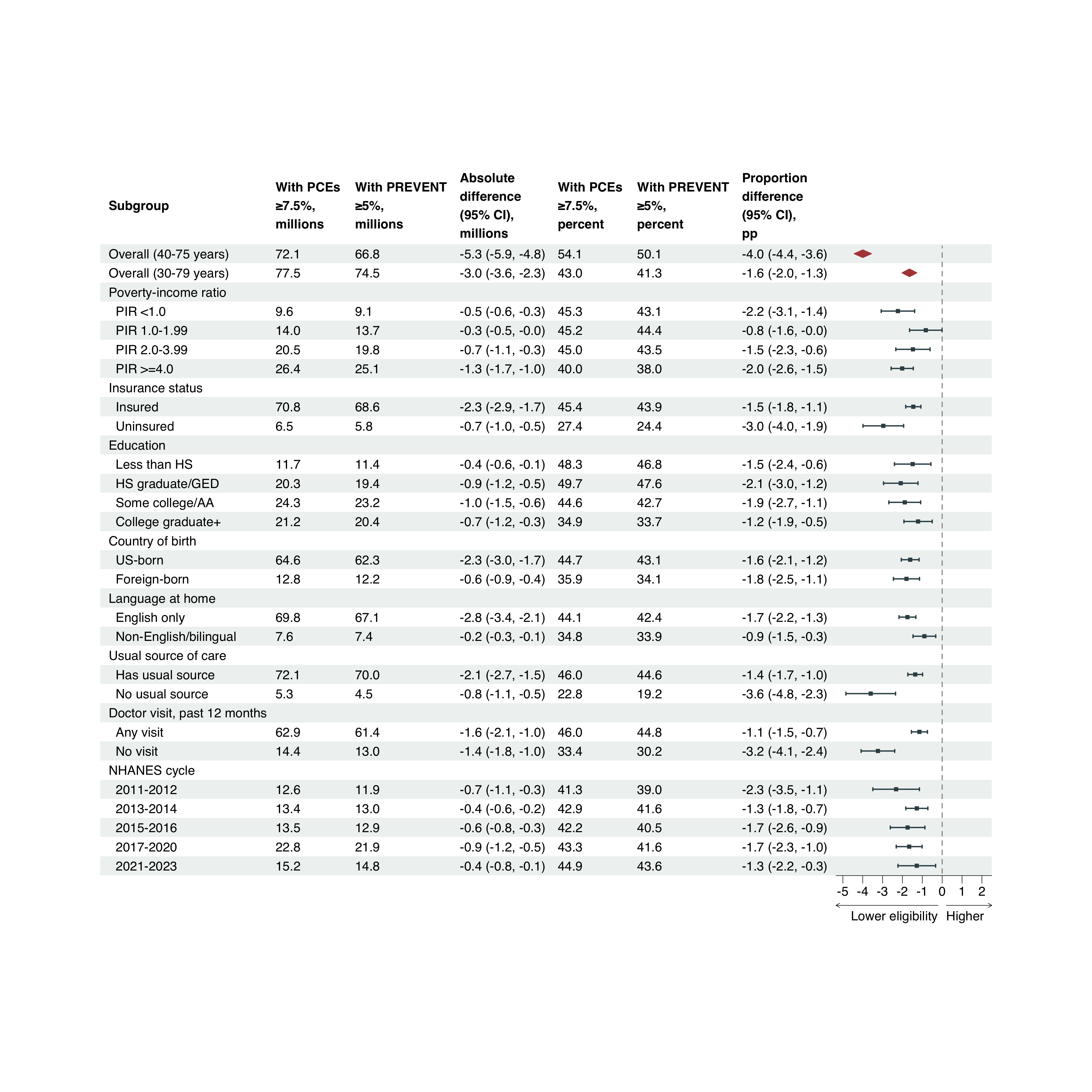}
\label{fig:sfig_ses}
\end{figure*}

\begin{figure*}[htbp]
\centering
\caption{\textbf{Multivariable predictors of change in statin recommendations under 2026 guidelines, US adults aged 40--75 years, NHANES 2011--2023.} Coefficients from a survey-weighted linear regression of the change in statin eligibility indicator ($+1$ = newly recommended, $0$ = no change, $-1$ = no longer recommended; class~1 threshold: PREVENT $\geq$5\% vs PCE $\geq$7.5\%) on demographic, clinical, and healthcare access covariates. Points represent regression coefficients; horizontal lines show 95\% CIs. Age is represented as categorical groups (50--59, 60--69, and 70--75 years vs reference 40--49 years); SBP is centered at 130 mmHg and scaled per 10 mmHg; LDL-C is centered at 120 mg/dL and scaled per 10 mg/dL; eGFR is centered at 90 mL/min/1.73m$^2$ and scaled per 10 units. Race/ethnicity coefficients are relative to non-Hispanic White.}
\includegraphics[width=\textwidth]{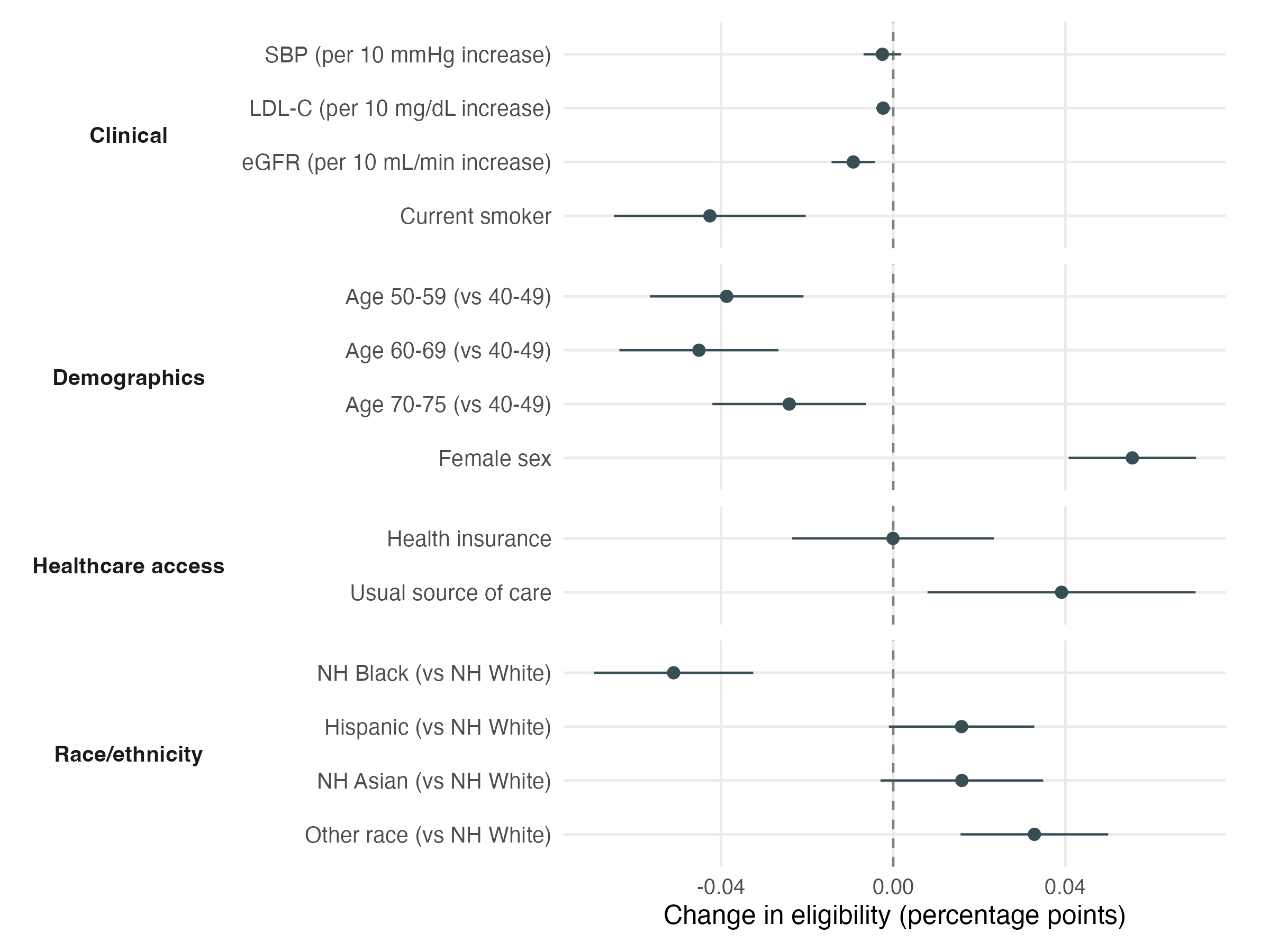}
\label{fig:sfig_regression}
\end{figure*}

\begin{figure*}[htbp]
\centering
\caption{\textbf{Scatterplot of PCE vs PREVENT 10-year ASCVD risk estimates among US adults aged 40--75 years without clinical ASCVD, NHANES 2011--2023.} Each point represents a survey participant. Vertical and horizontal dashed lines indicate relevant treatment thresholds. Points below the diagonal line have lower PREVENT than PCE risk estimates.}
\includegraphics[width=\textwidth]{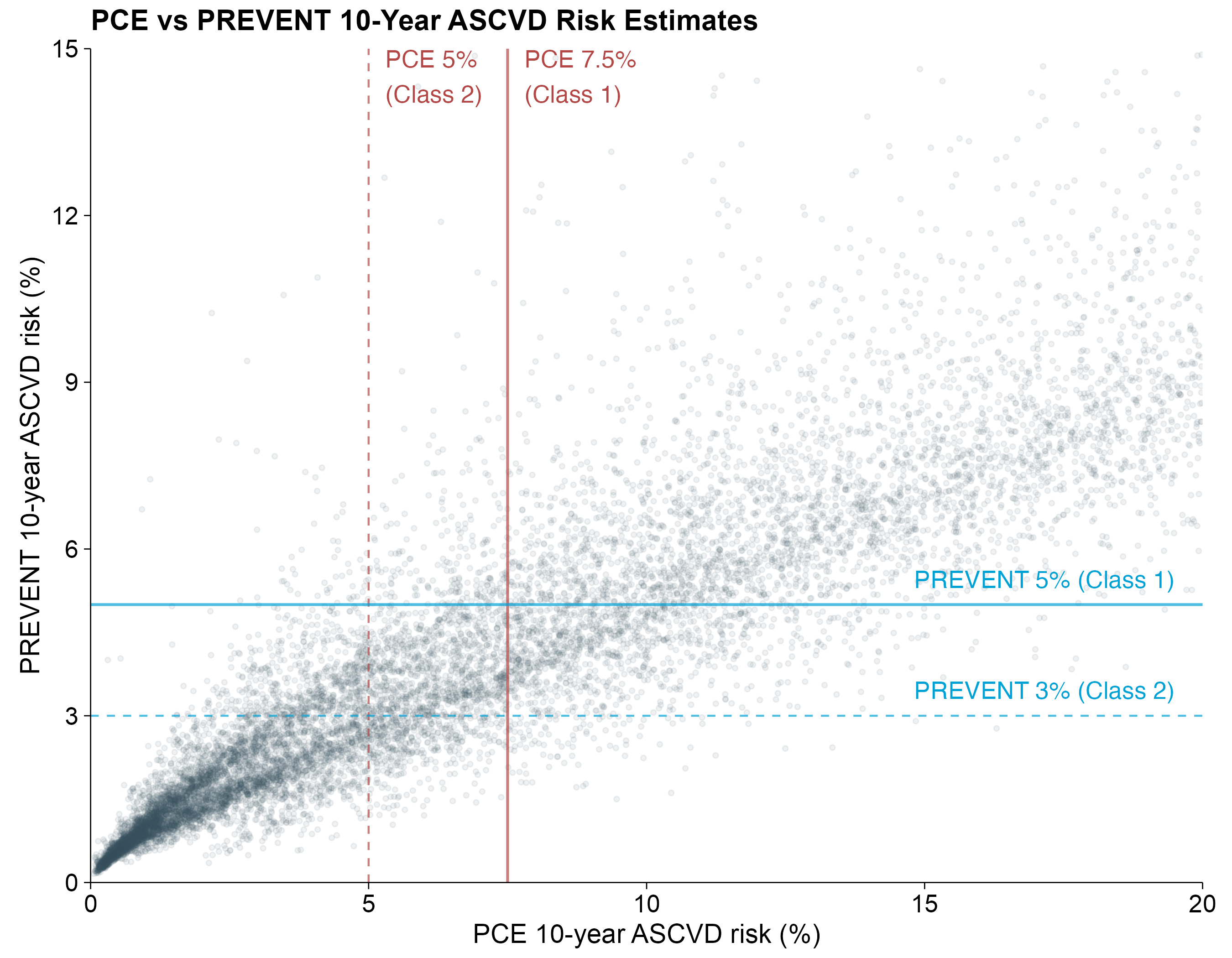}
\label{fig:sfig_scatter}
\end{figure*}

\begin{figure*}[htbp]
\centering
\caption{\textbf{Statin recommendations by indication type and age group under 2018 and 2026 guidelines, US adults aged 30--79 years, NHANES 2011--2023 (proportions).} Same as Figure~\ref{fig:figure2} but with the y-axis expressed as the weighted proportion of US adults rather than absolute counts. *Predicted risk based on PCE cannot be used to determine statin recommendations for these age groups because the PCEs are not validated for ages beyond 40--75 years.}
\includegraphics[width=\textwidth]{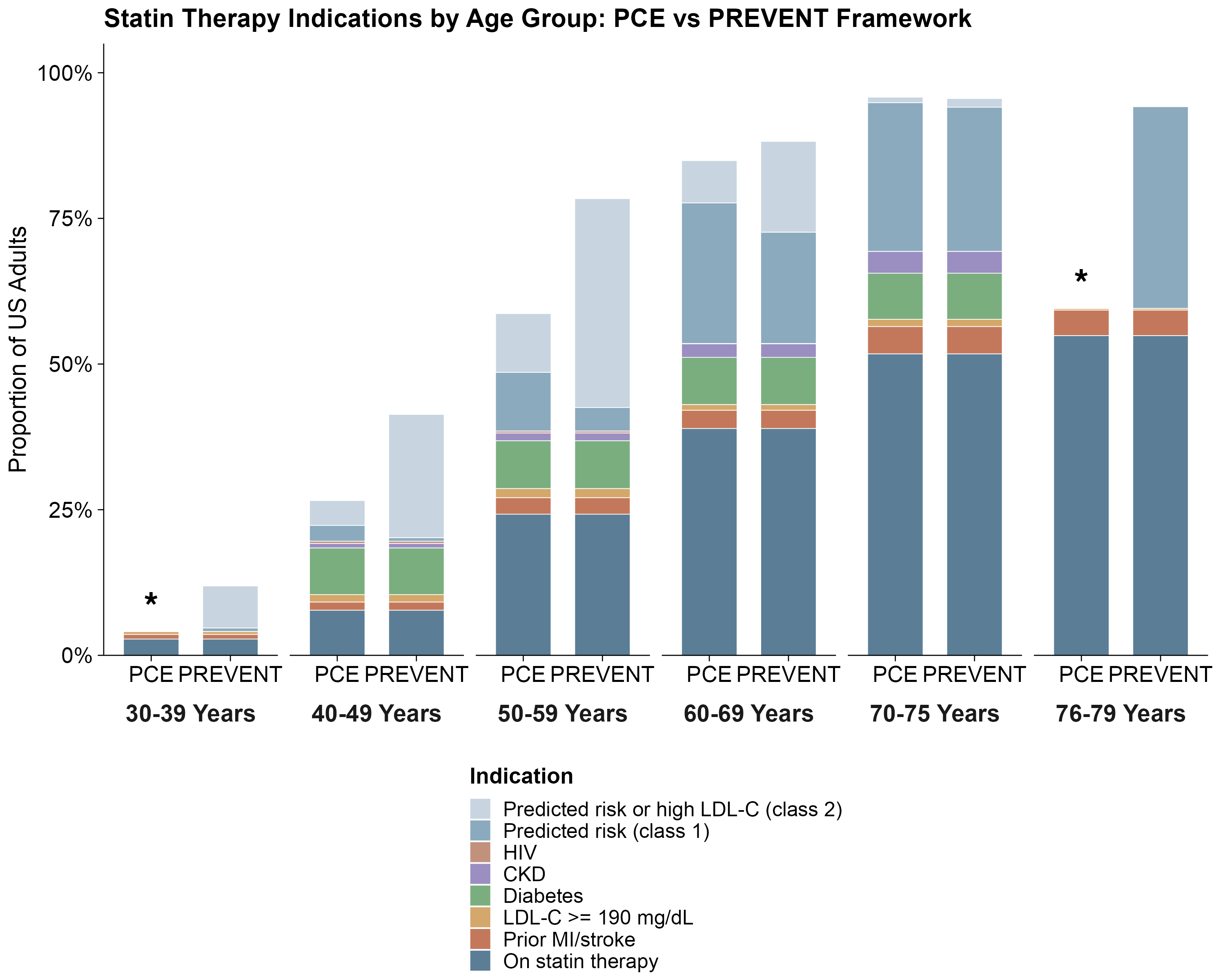}
\label{fig:sfig_prop_barplot}
\end{figure*}

\begin{figure*}[htbp]
\centering
\caption{\textbf{Sensitivity analysis (class~1 threshold): detreatment adjustment.} Net change in statin recommendations at the class~1 threshold (PREVENT $\geq$5\% vs PCE $\geq$7.5\%) after adjusting cholesterol values among participants receiving lipid-lowering therapy (statins, ezetimibe, PCSK9 inhibitors, fibrates, niacin) to approximate untreated baseline levels using published multiplicative correction factors. The 95\% CIs use the same paired within-person contrast approach as Figure~\ref{fig:figure3}. *The 76--79 age group is marked with an asterisk in the forest panel because the proportion difference falls far outside the range of other subgroups.}
\includegraphics[width=\textwidth]{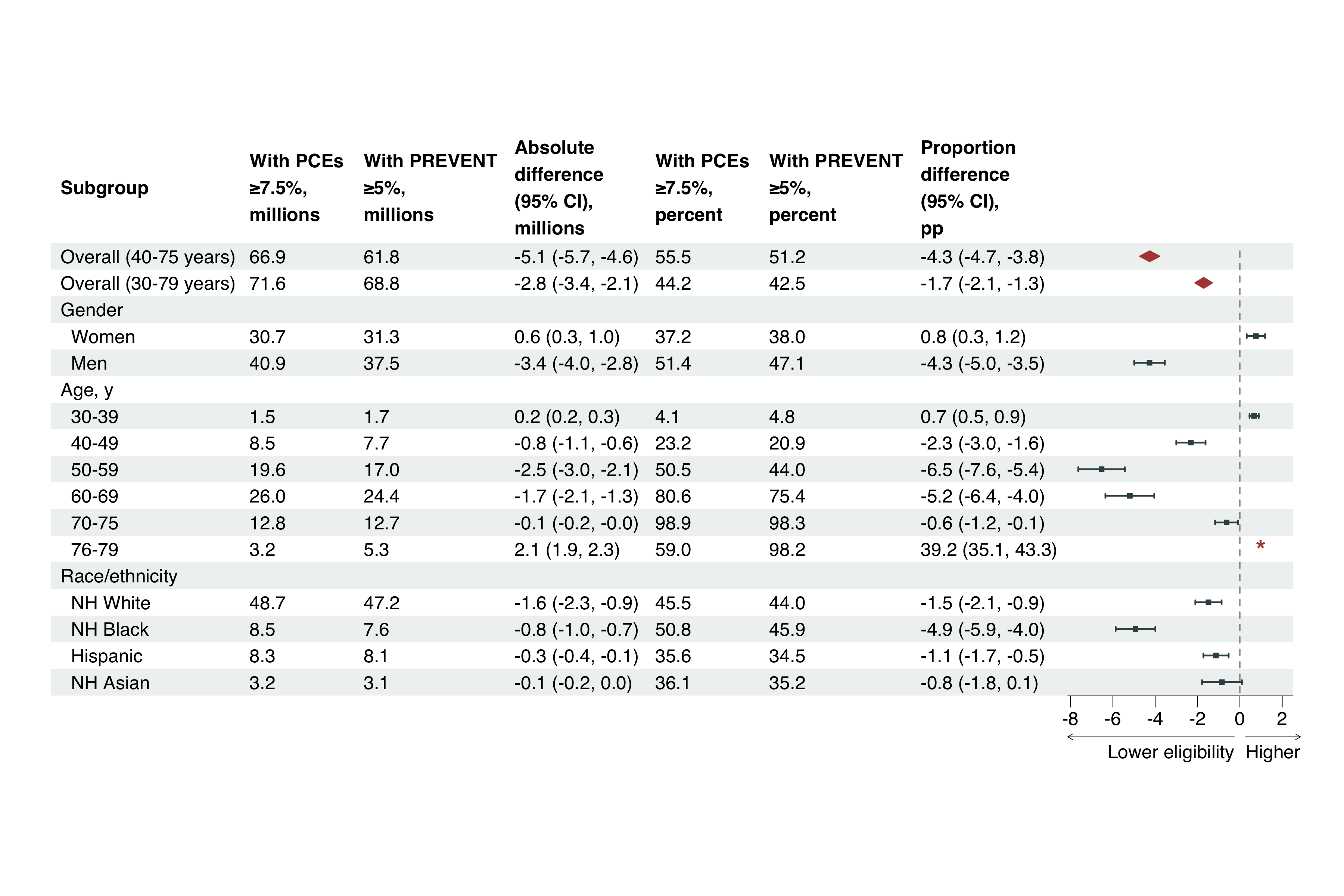}
\label{fig:sfig_detreat}
\end{figure*}

\begin{figure*}[htbp]
\centering
\caption{\textbf{Sensitivity analysis (class~1 threshold): excluding current statin users.} Net change in statin recommendations at the class~1 threshold (PREVENT $\geq$5\% vs PCE $\geq$7.5\%) restricted to adults not currently receiving statin therapy (n = 18{,}446), to isolate the change in new recommendations among untreated adults. The 95\% CIs use the same paired within-person contrast approach as Figure~\ref{fig:figure3}. *The 76--79 age group is marked with an asterisk in the forest panel because the proportion difference falls far outside the range of other subgroups.}
\includegraphics[width=\textwidth]{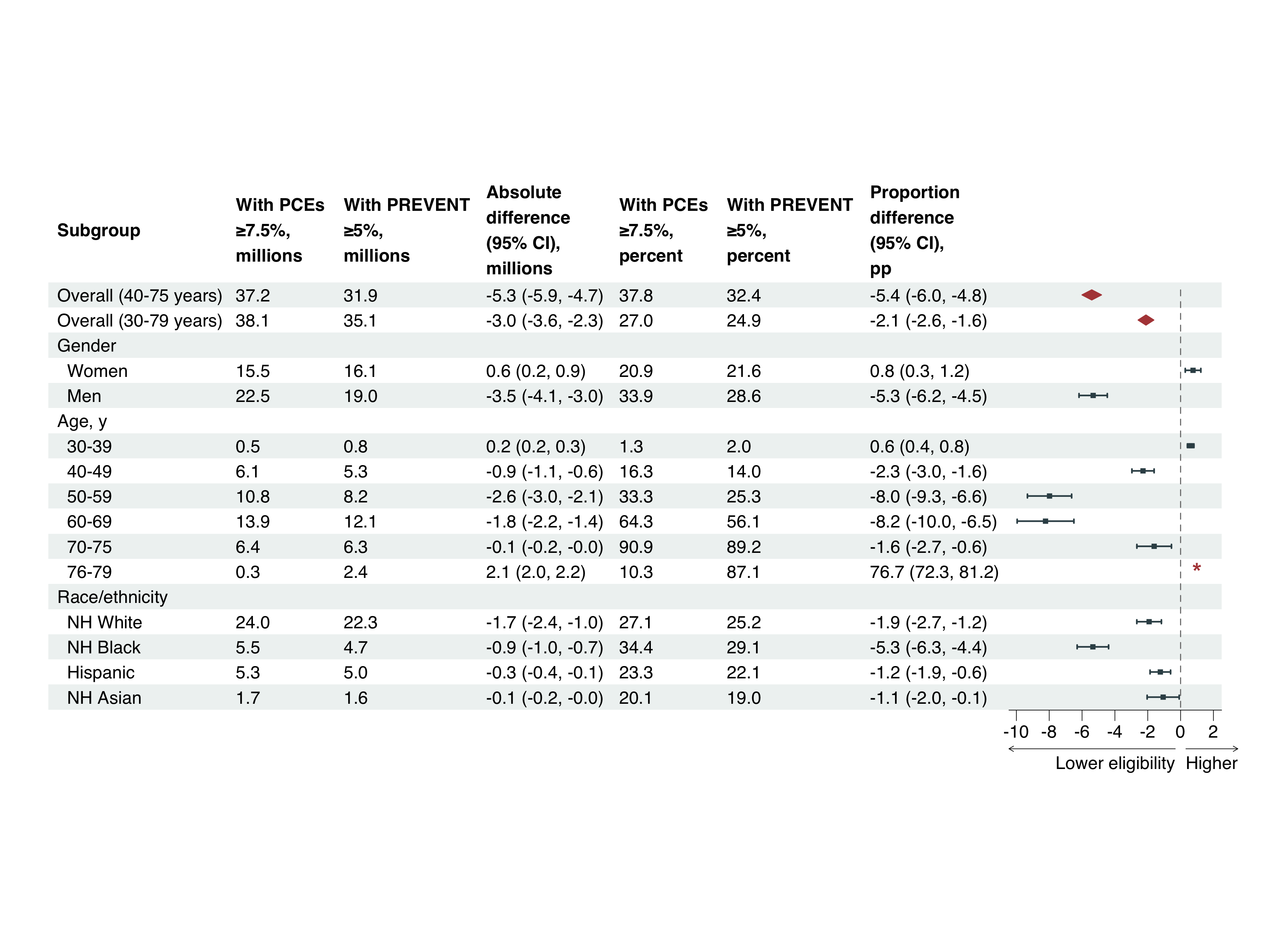}
\label{fig:sfig_nostatin}
\end{figure*}

\begin{figure*}[htbp]
\centering
\caption{\textbf{Sensitivity analysis (class~1 threshold): in-range equation inputs only.} Net change in statin recommendations at the class~1 threshold (PREVENT $\geq$5\% vs PCE $\geq$7.5\%) restricted to participants with all risk factor values within the recommended input ranges for both PCE and PREVENT. The 95\% CIs use the same paired within-person contrast approach as Figure~\ref{fig:figure3}. *The 76--79 age group is marked with an asterisk in the forest panel because the proportion difference falls far outside the range of other subgroups.}
\includegraphics[width=\textwidth]{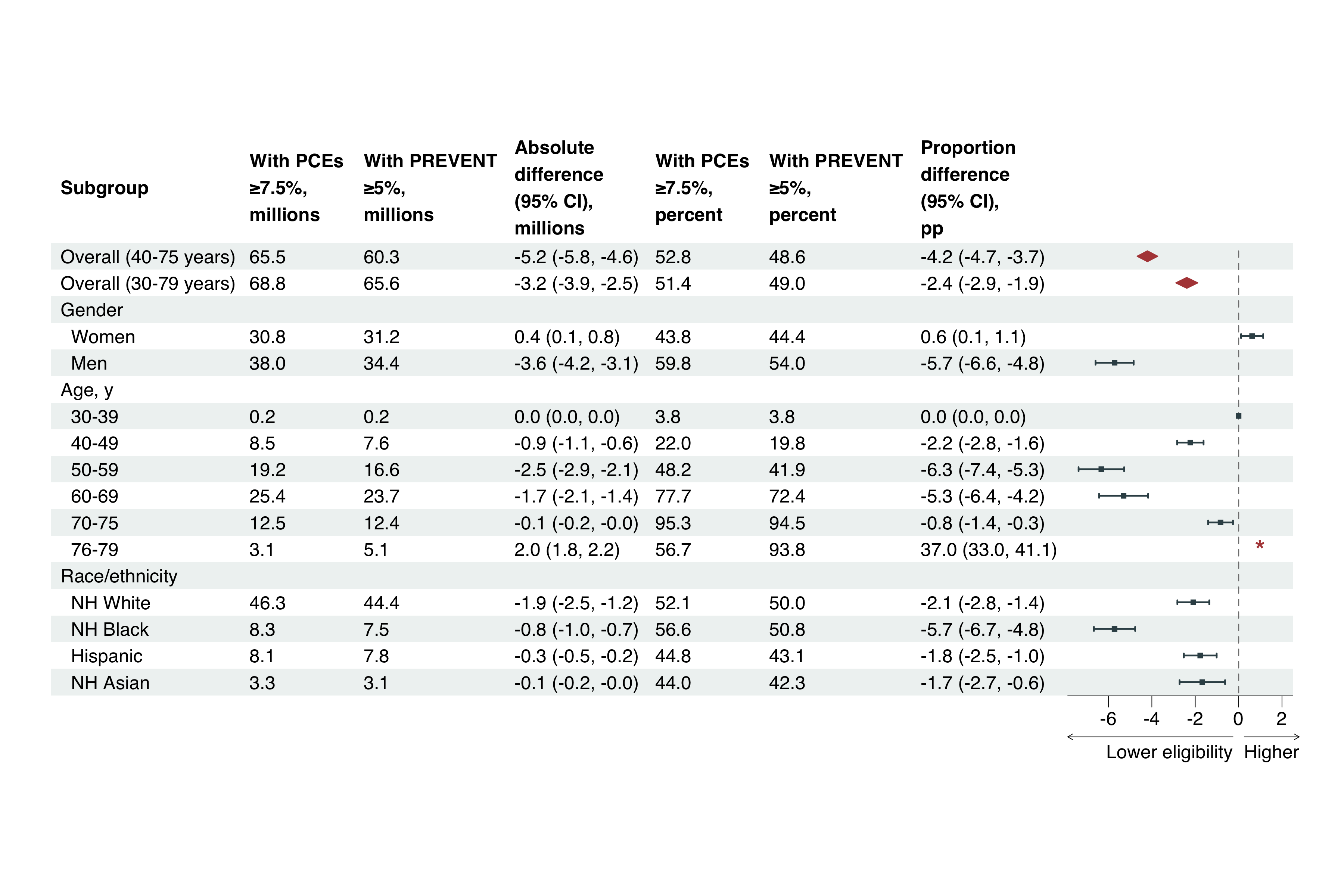}
\label{fig:sfig_inrange}
\end{figure*}

\begin{figure*}[htbp]
\centering
\caption{\textbf{Sensitivity analysis (class~1 threshold): fasting subsample with fasting-subsample weights.} Net change in statin recommendations at the class~1 threshold (PREVENT $\geq$5\% vs PCE $\geq$7.5\%) restricted to the NHANES fasting subsample (n = 10{,}762) using fasting-subsample survey weights. The 95\% CIs use the same paired within-person contrast approach as Figure~\ref{fig:figure3}. *The 76--79 age group is marked with an asterisk in the forest panel because the proportion difference falls far outside the range of other subgroups.}
\includegraphics[width=\textwidth]{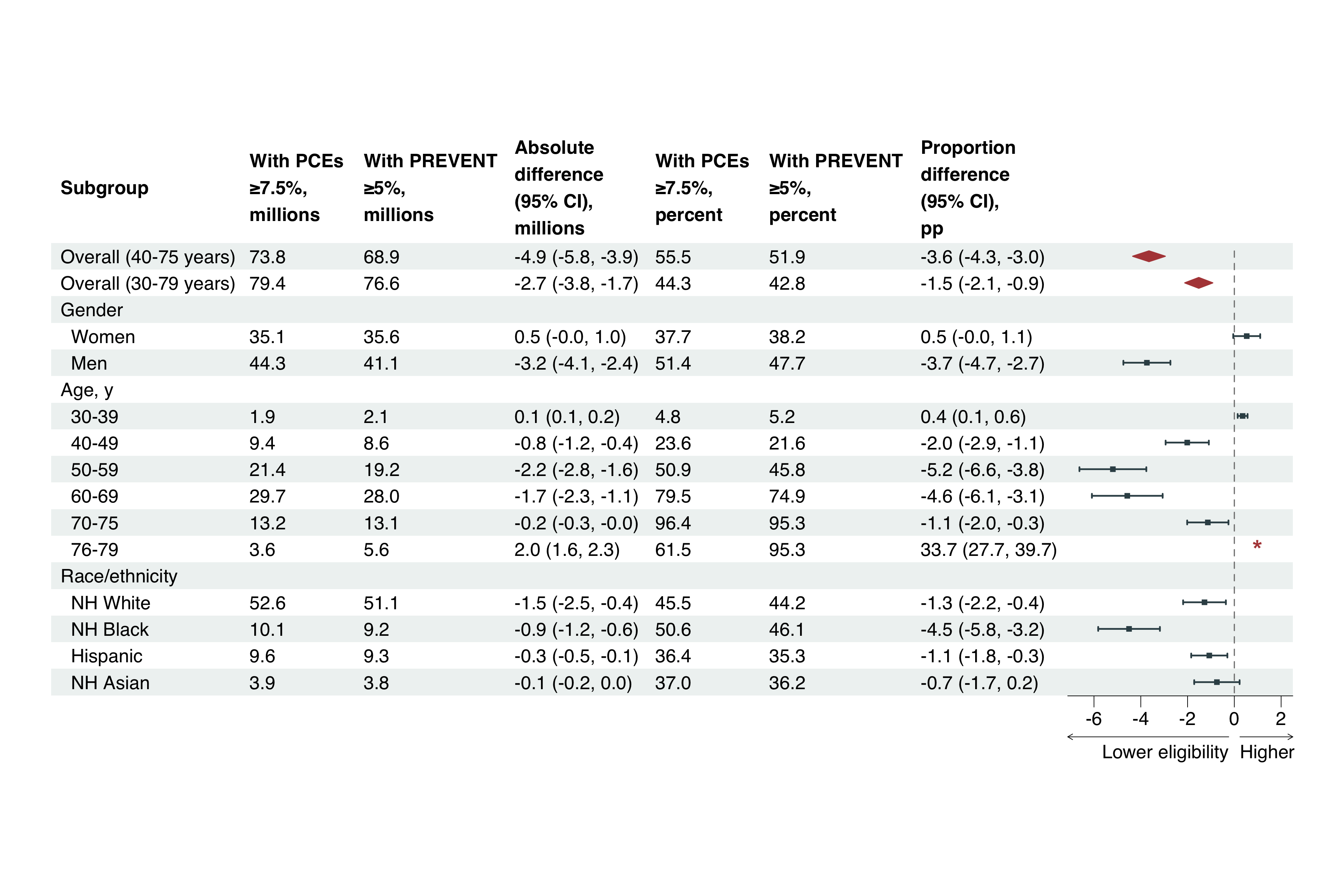}
\label{fig:sfig_fasting}
\end{figure*}


\begin{figure*}[htbp]
\centering
\caption{\textbf{Subgroup forest plot: net change in statin recommendations by socioeconomic and NHANES cycle subgroups, class~2 threshold (PREVENT $\geq$3\%+ vs PCE $\geq$5\%), US adults aged 30--79 years, NHANES 2011--2023.} Same subgroups and methods as Figure~\ref{fig:sfig_ses}, but evaluated at the class~2 threshold, which additionally recommends statins for adults aged 30--59 years with LDL-C 160--189 mg/dL or 30-year PREVENT ASCVD risk $\geq$10\%. Positive values indicate increased statin recommendations under 2026 guidelines. The 95\% CIs use the same paired within-person contrast approach as Figure~\ref{fig:figure3}.}
\includegraphics[width=\textwidth]{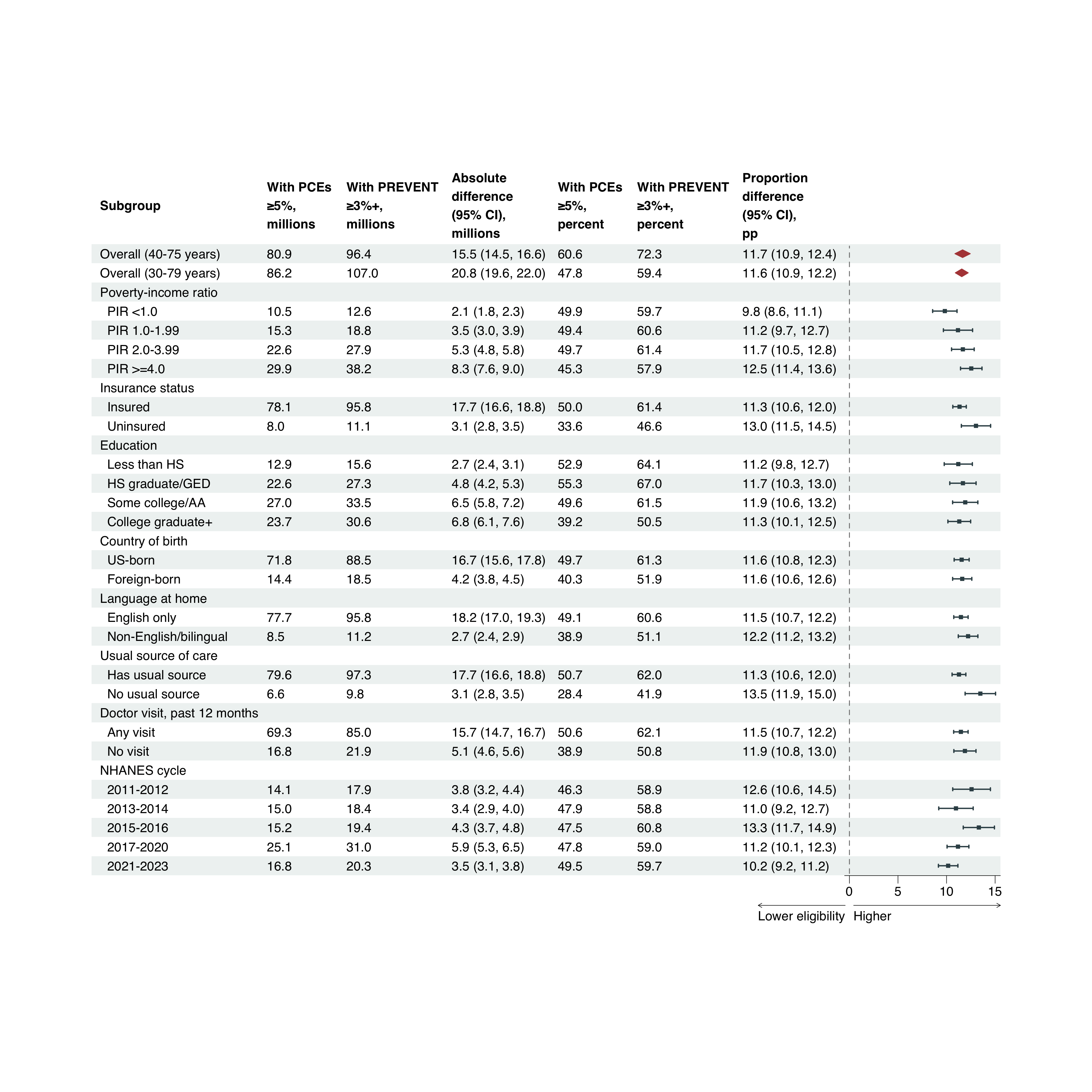}
\label{fig:sfig_ses_c2}
\end{figure*}

\begin{figure*}[htbp]
\centering
\caption{\textbf{Multivariable predictors of change in statin recommendations under 2026 guidelines at the class~2 threshold, US adults aged 40--75 years, NHANES 2011--2023.} Same model specification as Figure~\ref{fig:sfig_regression}, but with the outcome defined as the change in statin eligibility at the class~2 threshold (PREVENT $\geq$3\%+ vs PCE $\geq$5\%). Points represent regression coefficients; horizontal lines show 95\% CIs. Notably, the age 50--59 coefficient is strongly positive (reflecting the 30-year risk criterion), while older age groups show negative coefficients (the 30-year risk criterion does not apply above age 59). Female sex also shows a large positive coefficient.}
\includegraphics[width=\textwidth]{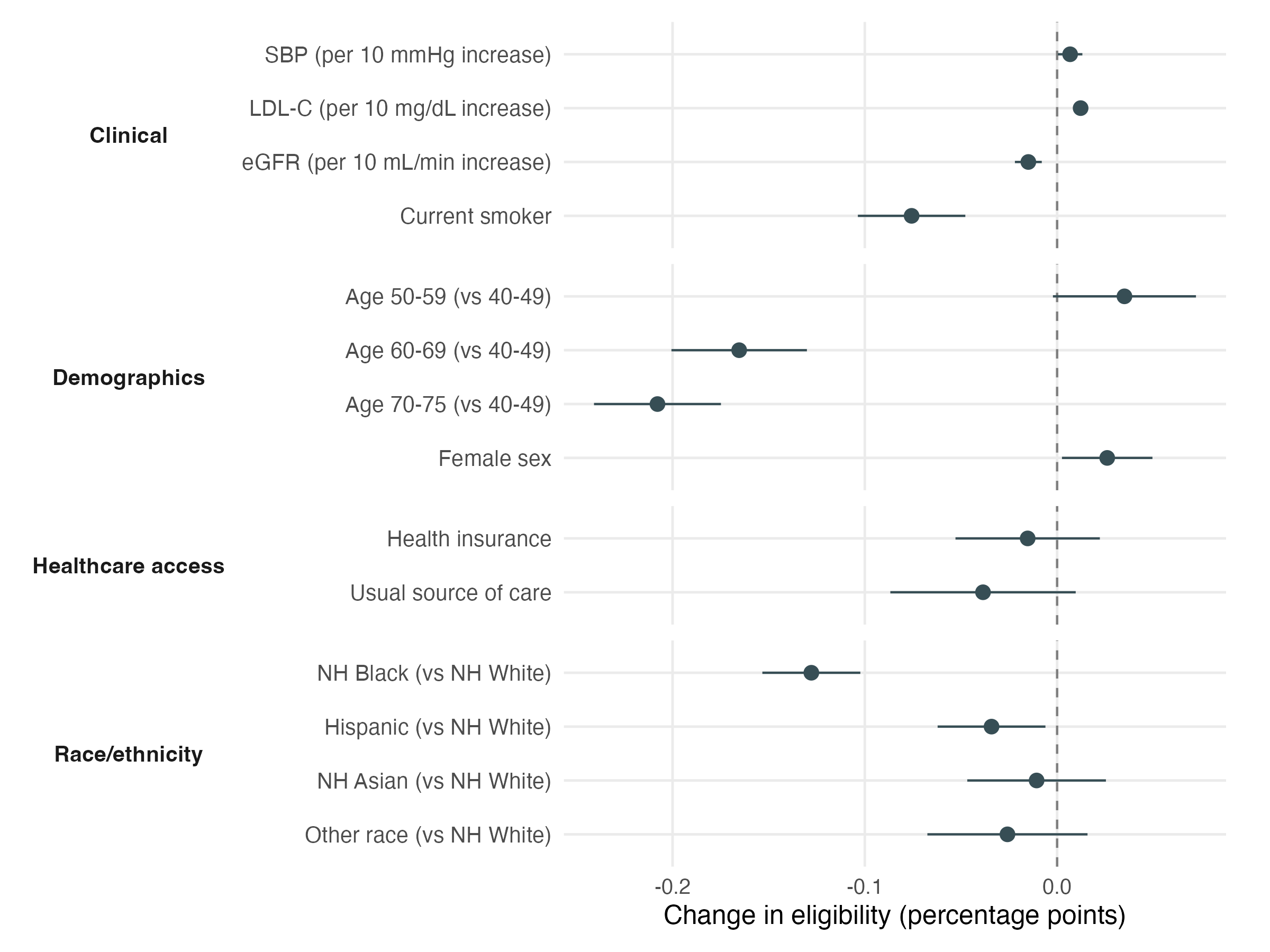}
\label{fig:sfig_regression_c2}
\end{figure*}

\begin{figure*}[htbp]
\centering
\caption{\textbf{Sensitivity analysis: detreatment adjustment, class~2 threshold.} Net change in statin recommendations at the class~2 threshold (PREVENT $\geq$3\%+ vs PCE $\geq$5\%) after adjusting cholesterol values among participants receiving lipid-lowering therapy to approximate untreated baseline levels. The 95\% CIs use the same paired within-person contrast approach as Figure~\ref{fig:figure3}. *The 76--79 age group is marked with an asterisk in the forest panel because the proportion difference falls far outside the range of other subgroups.}
\includegraphics[width=\textwidth]{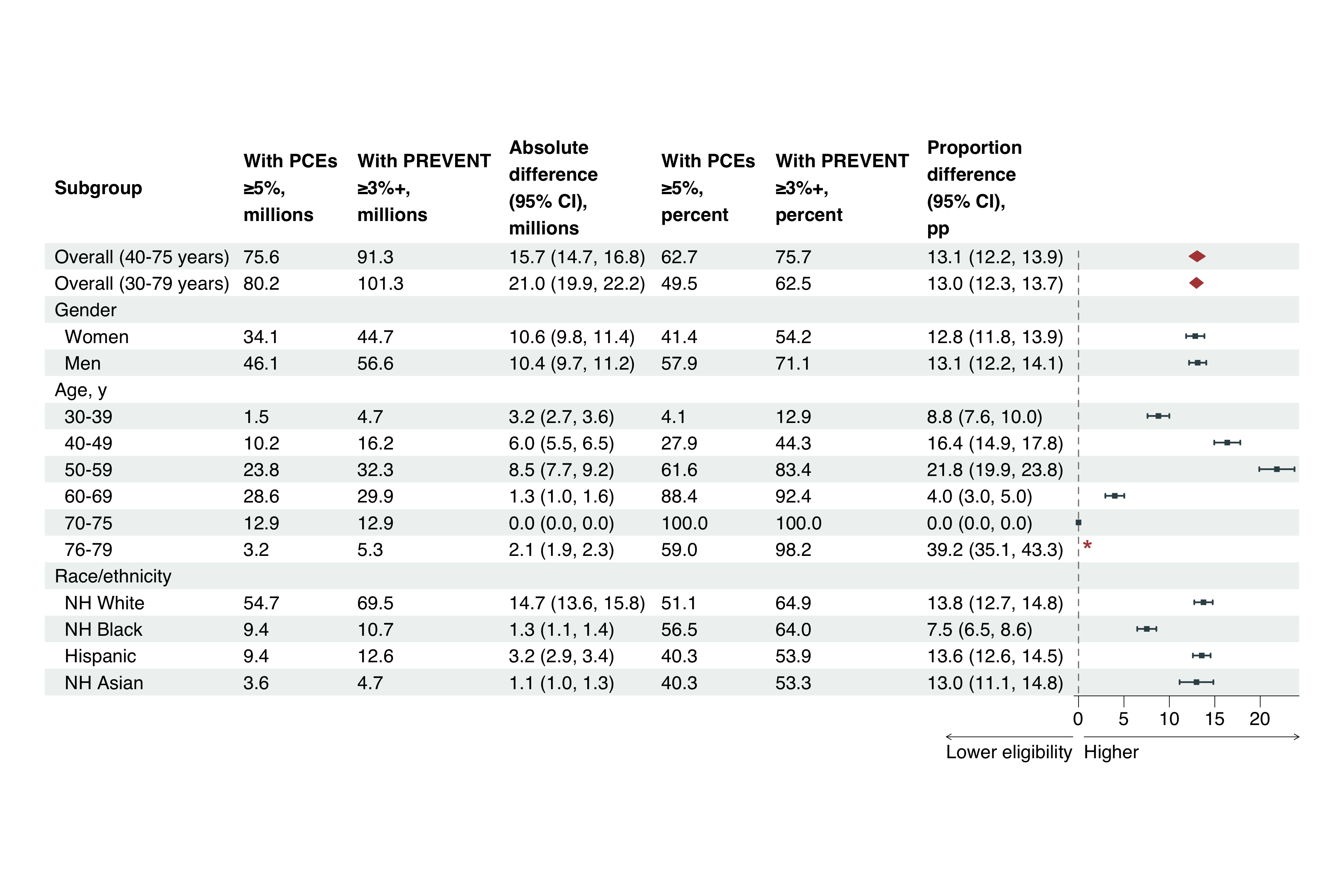}
\label{fig:sfig_detreat_c2}
\end{figure*}

\begin{figure*}[htbp]
\centering
\caption{\textbf{Sensitivity analysis: excluding current statin users, class~2 threshold.} Net change in statin recommendations at the class~2 threshold (PREVENT $\geq$3\%+ vs PCE $\geq$5\%) restricted to adults not currently receiving statin therapy, to isolate the change in new recommendations among untreated adults. The 95\% CIs use the same paired within-person contrast approach as Figure~\ref{fig:figure3}. *The 76--79 age group is marked with an asterisk in the forest panel because the proportion difference falls far outside the range of other subgroups.}
\includegraphics[width=\textwidth]{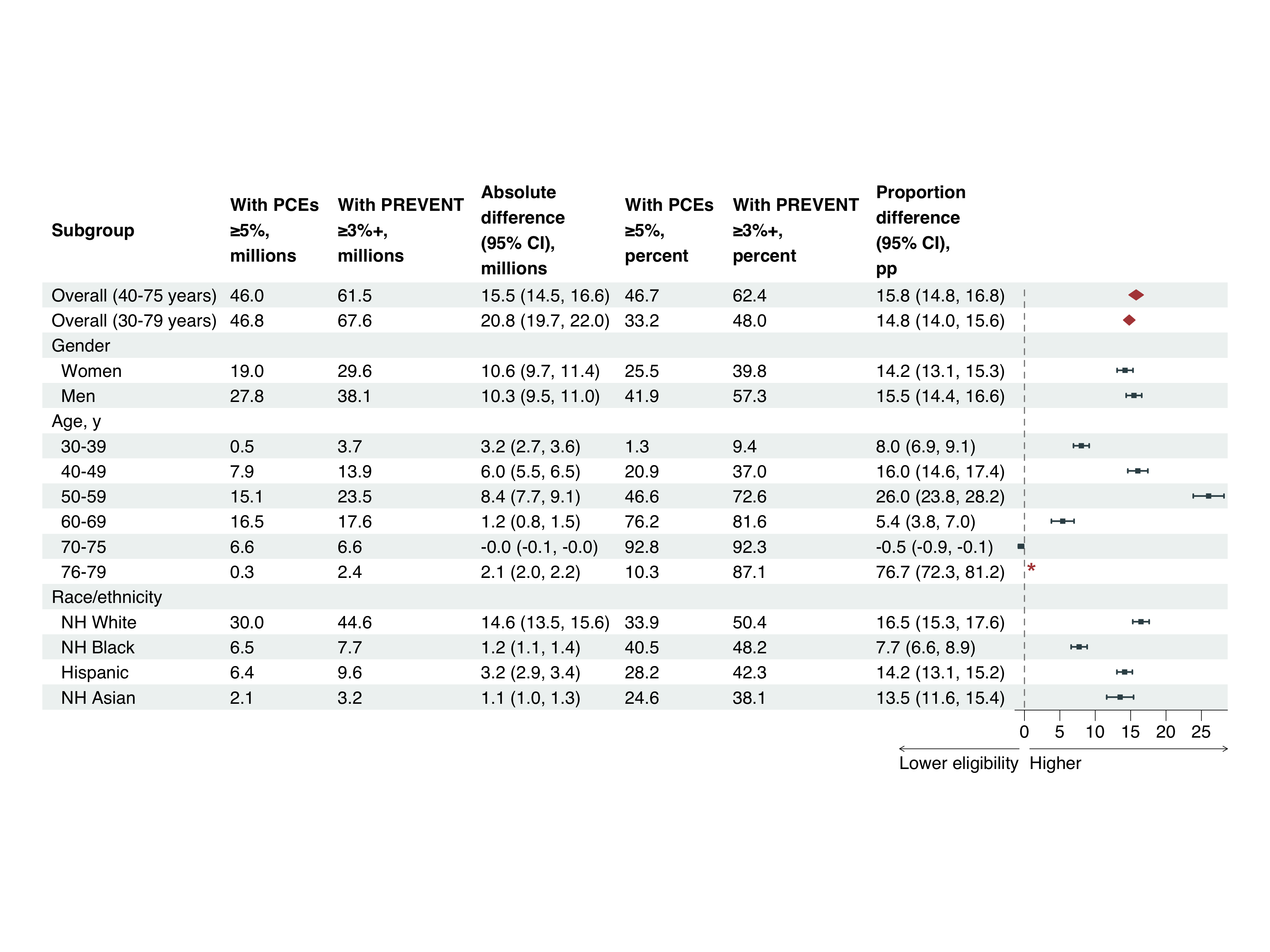}
\label{fig:sfig_nostatin_c2}
\end{figure*}

\begin{figure*}[htbp]
\centering
\caption{\textbf{Sensitivity analysis: in-range equation inputs only, class~2 threshold.} Net change in statin recommendations at the class~2 threshold (PREVENT $\geq$3\%+ vs PCE $\geq$5\%) restricted to participants with all risk factor values within the recommended input ranges for both PCE and PREVENT. The 95\% CIs use the same paired within-person contrast approach as Figure~\ref{fig:figure3}. *The 76--79 age group is marked with an asterisk in the forest panel because the proportion difference falls far outside the range of other subgroups.}
\includegraphics[width=\textwidth]{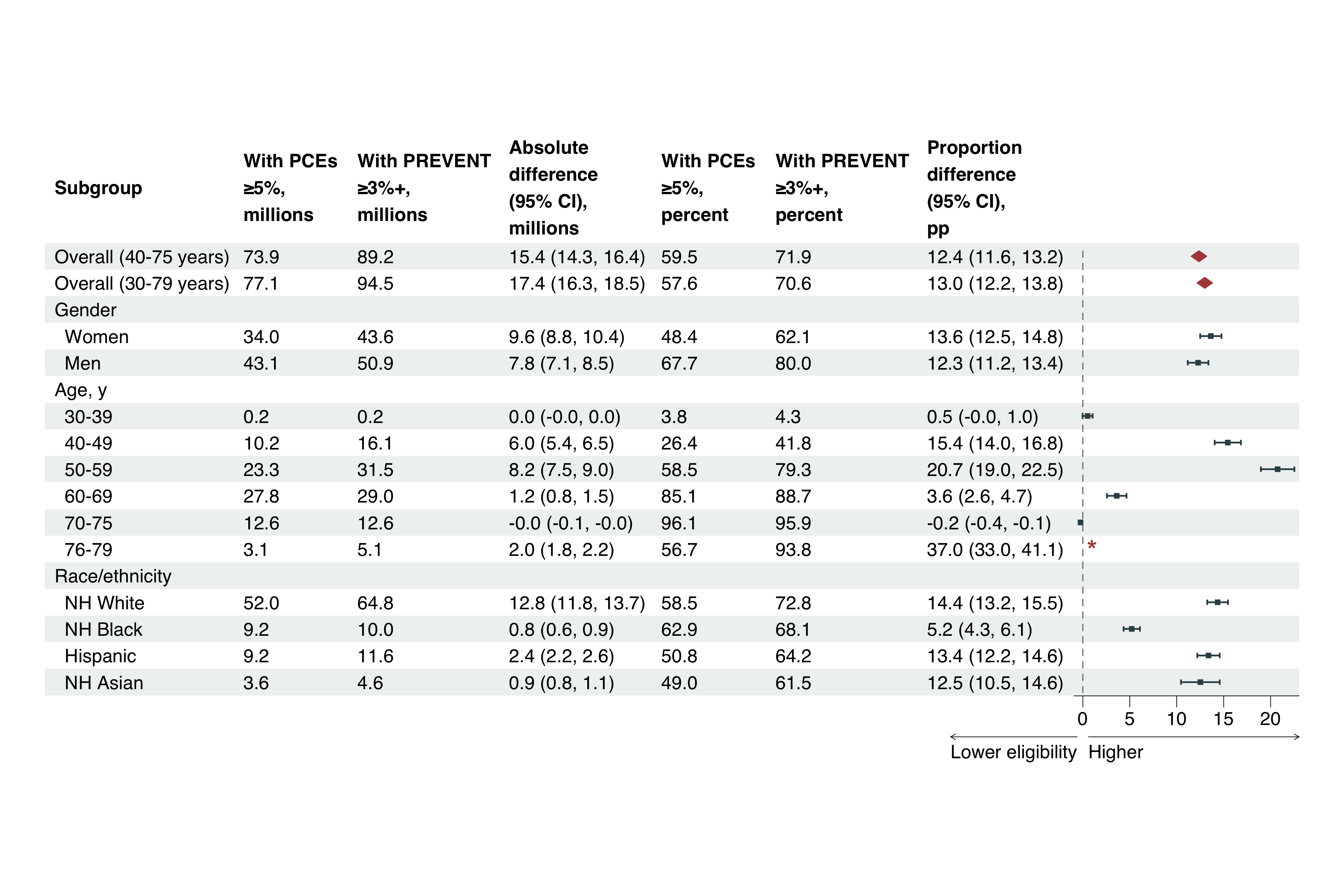}
\label{fig:sfig_inrange_c2}
\end{figure*}

\begin{figure*}[htbp]
\centering
\caption{\textbf{Sensitivity analysis: fasting subsample with fasting-subsample weights, class~2 threshold.} Net change in statin recommendations at the class~2 threshold (PREVENT $\geq$3\%+ vs PCE $\geq$5\%) restricted to the NHANES fasting subsample using fasting-subsample survey weights. The 95\% CIs use the same paired within-person contrast approach as Figure~\ref{fig:figure3}. *The 76--79 age group is marked with an asterisk in the forest panel because the proportion difference falls far outside the range of other subgroups.}
\includegraphics[width=\textwidth]{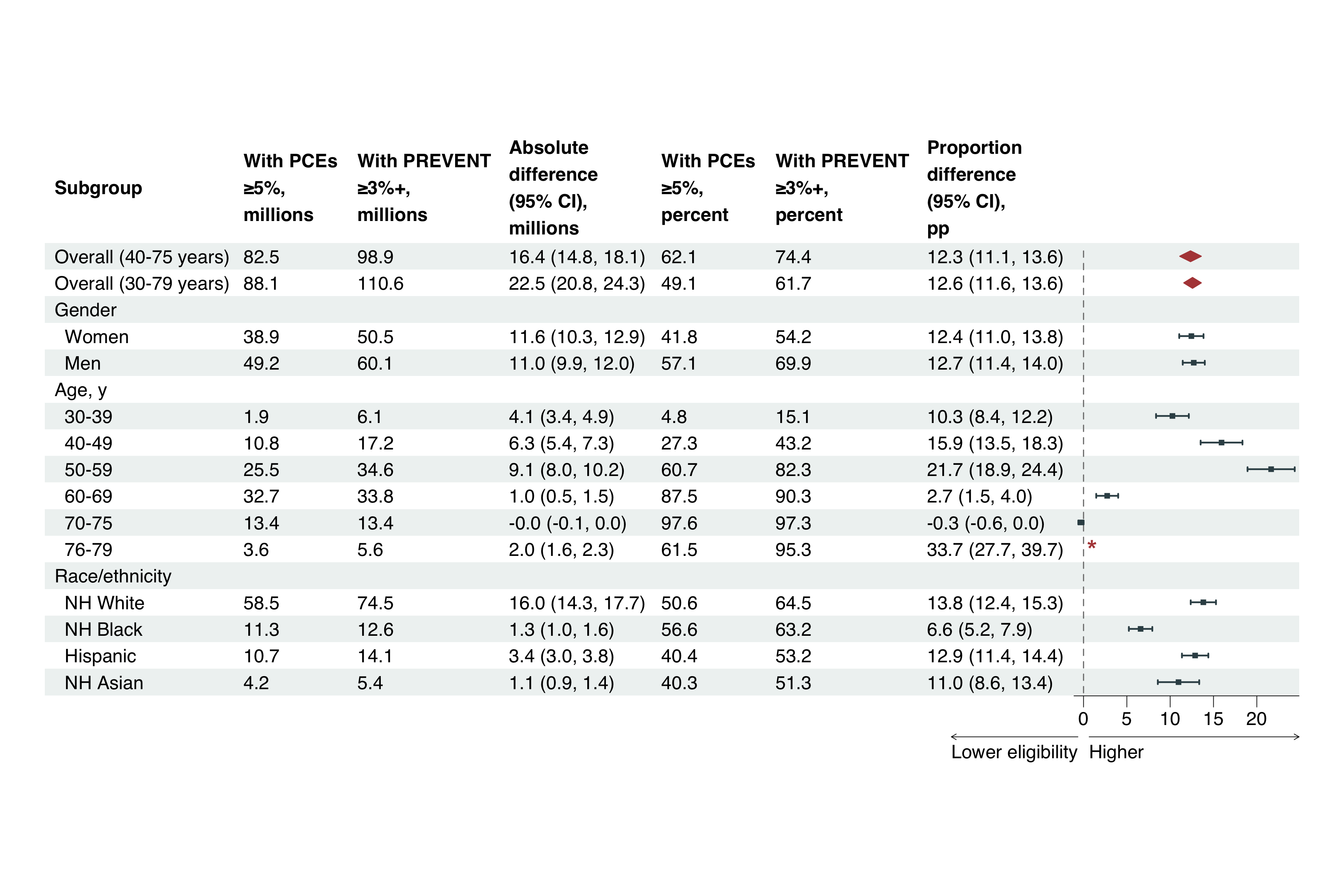}
\label{fig:sfig_fasting_c2}
\end{figure*}

\end{document}